\documentclass{aa}  

\usepackage{graphicx}
\usepackage{txfonts}
\usepackage{multirow}
\usepackage{xcolor}
\usepackage{ulem}
\usepackage{comment}
\usepackage{natbib}
\usepackage{amssymb}
\usepackage{scrextend} %footnotes
\usepackage{placeins} %appendix
\usepackage{hyperref}
\usepackage{caption}
\usepackage{subfigure}
\usepackage{soul}
\hypersetup{
    colorlinks=true,
    linkcolor=blue,
    filecolor=magenta,      
    urlcolor=blue,
    citecolor=blue,
    } 

\begin{document}

\title{Radio study of the colliding wind binary HD~93129A near periastron and its surroundings}

   \author{P. Benaglia\inst{1}
          \and
          S. del Palacio\inst{2}
          \and
          J. Saponara\inst{1}
          \and
          A.~B. Blanco\inst{3}
          \and
          M. De Becker\inst{3}
          \and
          B. Marcote\inst{4,5}
          }

   \institute{Instituto Argentino de Radioastronomía, CONICET-CICPBA-UNLP, CC5 (1897) Villa Elisa, Prov. de Buenos Aires, Argentina\\
             \email{paula@iar-conicet.gov.ar}
         \and
             Department of Space, Earth and Environment, Chalmers University of Technology, SE-412 96 Gothenburg, Sweden
         \and
            Space Sciences, Technologies and Astrophysics Research (STAR) Institute, University of Liège, Quartier Agora, 19c, Allée du 6 Août, B5c, 4000, Sart Tilman, Belgium
        \and
        Joint Institute for VLBI ERIC, Oude Hoogeveensedijk 4, 7991 PD, Dwingeloo, The Netherlands
       \and
        ASTRON, Netherlands Institute for Radio Astronomy, Oude Hoogeveensedijk 4, 7991 PD, Dwingeloo, The Netherlands
    }

   \date{Received ...; accepted ...  }

% \abstract{}{}{}{}{}
% 5 {} token are mandatory
 
  \abstract
   {HD~93129A is an O+O stellar system whose colliding-wind region (CWR) has been mapped by high angular resolution observations at cm wavelengths. The synchrotron nature of the radio emission confirms its particle accelerator status. According to the analysis of astrometric measurements since 1996, the system has an orbital period of $\sim$120 yr and recently went through its periastron passage.}
   {We attempted to characterise the radio emission during  the epoch of periastron passage, when the particle density and the local magnetic energy density in the CWR increase.} 
   {We monitored HD~93129A and its surroundings in bands centred at 2.1, 5.5, and 9.0~GHz, with the Australia Telescope Compact Array (ATCA) over a time span of 17 months, with an approximate 2-month cadence. Previous ATCA data at similar frequencies and data collected using other radio observatories were also included.}
   {We obtained radio light curves in subbands per band per epoch.  The flux densities show an average rise of a factor four from 2003 to {2018, with the caveat that the 2009–2018 time lapse is devoid of data}, and a similar decay between {2018 to 2020. We fit the spectral energy distribution of quasi-simultaneous data at three epochs and find that the magnetic-to-thermal pressure ratio $\eta_B$ does not remain constant along the orbit, possibly suggesting magnetic field amplification close to periastron. In the 2019 epoch, we estimate a magnetic field strength of $\approx$1.1~G in the apex of the CWR (corresponding to $\eta_B \approx 0.085$). 
   The evolution of the SED and spectral index along several epochs in 2019--2020} is also presented. By combining ATCA and ASKAP images, a spectral index map was obtained in an area of 30$'$ size, including the surroundings of $\eta$\,Car, {revealing positive and negative spectral indices.} 
   The radio emission in the direction of other massive binary systems in the fields of view (WR~22, WR~25 and HD~93250) was measured and briefly discussed.}
   {Intensive radio monitoring of a CWB during key orbital phases allows us to track the evolution of physical conditions in the shocks. {The general trend of decreasing emission of HD~93129A in the high-frequency bands in 2019--2020 suggests that the system is at post-periastron, consistent with model predictions.}
}


   \maketitle
%
%________________________________________________________________

\section{Introduction}
\label{sec-introduction}

Massive, early-type stars, belonging to the OB and Wolf-Rayet classes, exhibit strong winds that transfer material and energy to their surroundings. A large proportion of these stars are part of binary or higher multiplicity systems \citep{Offner2023}. In such systems, if the individual stars are sufficiently close, their stellar winds interact in a colliding-wind region (CWR) and the systems are referred to as colliding-wind binaries (CWBs). Since the 1970s, these stars have been identified as sources of thermal and non-thermal (NT) radio emission. The former can be attributed to the free--free emission of extended envelopes \citep[e.g.,][]{wrightbarlow1975}, while the latter is due to synchrotron radiation \citep{White1985}. 
 
High Mach number shocks in CWBs constitute ideal sites for diffusive shock acceleration, leading to the acceleration of charged particles to relativistic velocities. The subset of CWBs showing evidence for particle acceleration is referred to as particle-accelerating colliding-wind binaries (PACWBs). These systems are revealed by the signature of NT emission processes \citep{eichlerusov1993,DeBecker2007}. The modelling of these processes lends support to the idea that such systems are capable to produce radiation from the radio domain to high energies \citep{pittard2006b,Reimer2006,Reitberger2017}. These processes occur in a manner analogous to those observed in supernova remnants, albeit with higher mass, photon, and magnetic energy densities. 

The most comprehensive catalogue of PACWB was published by \citet{pacwbc}\footnote{ \url{https://www.astro.uliege.be/~debecker/pacwb/}}. The updated online version of the catalogue lists 54 members. PACWBs offer an invaluable opportunity to study particle acceleration and non-thermal processes across a broad energy range. PACWBs at a few kpc can be detected in the radio domain, and only a few have had their intensity distribution mapped \citep[see for instance][this last hereafter B15, and references therein]{sanchez2019,Marcote2021,DeBecker2024,benaglia2015}. The same population of relativistic electrons involved in synchrotron radio emission has the potential to give rise to high-energy emission. In particular, these objects are expected to be non-thermal X-ray emitters through inverse Compton scattering. To date, two systems have been identified as NT X-ray emitters: $\eta$\,Car \citep[HD~98308,][]{hamaguchi2018} and Apep \citep{delPalacio2023}. The latter is unique as it is the only PACWB identified as a NT emitter both in the radio domain and in X-rays. We can also count a possible detection of NT X-rays in HD~93129A \citep{delPalacio2020}. $\eta$\,Car is also the only CWB identified as a NT X-ray and $\gamma$-ray emitter, {as per \citet{tavani2009} with \textit{AGILE}, \citet{abdo2010} with \textit{Fermi} and \citet{hesscoll2025}.
And} another widely studied one, WR~11 (also known as $\gamma^2$ Velorum) is associated with a GeV source detected by the \textit{Fermi} satellite \citep[e.g.][]{fermi3}. The $\gamma$-ray emission is very likely due to a population of relativistic protons {\citep[see][]{white2020,martidevesa2020}}.

One of the most instructive parts of the orbit for observing the emission from a CWB is in the vicinity of periastron passage. At that epoch, the physical parameters governing the wind interaction change quickly and coincide with a significant rise of the particle density in the CWR {and its direct vicinity in the winds. This causes the synchrotron emission region to be more exposed to free--free absorption (FFA). This is especially well illustrated for instance in the case of WR\,140 \citep{dougherty2005}, where the synchrotron emission becomes dimmer as the system gets closer to periastron.}
In parallel, that specific region is exposed to a stronger radiation field from the stellar photospheres and the local magnetic energy density increases as well. This results in remarkable manifestations, especially in the high-energy spectral range, including a significant rise of the intrinsic X-ray emission from the colliding winds and a rise of the synchrotron radio emission. Monitoring the radio flux of such systems around the periastron passage provides thus essential information for characterising them, which in turn can be used to predict or explain high-energy emission from the CWR. {For a comprehensive overview of the non-thermal emission processes at work in such systems, we refer to the theoretical paper by \citet{pittard2006b}.}

Among PACWBs, the HD~93129A system located in the Car\,OB1 association contains one of the earliest, hottest, and most luminous stars in the Galaxy. It is currently close to periastron passage, with a predicted occurrence between 2017 and 2025 \citep[B15 and][]{maiz2017}, with an orbital period of about 120 years. Furthermore, Car OB1 harbors additional massive binary systems in the vicinity of HD~93129A, among which the most noticeable is $\eta$\,Car.
Due to HD~93129A's low declination, until very recently,
low frequency radio interferometric observations of the system were only feasible from Australian facilities. Consequently, we conducted an observing campaign of HD~93129A during 2019--2020 with the Australia Telescope Compact Array (ATCA). 

The present paper summarises the results of the mentioned campaign. The monitoring was designed to characterise its radio emission close to the most informative and critical orbital phase for investigating the physics of its colliding winds. Section\,\ref{secttarget} presents the relevant characteristics of the stellar system and its environment. Section\,\ref{sectobs} describes the compilation of the observations and the data reduction process. Section\,\ref{sectresults} reports on the results, and Sect.\,\ref{sectdisc} and Sect.\,\ref{sectdiscother} on their analysis and discussion. Our conclusions are given in Sect.\,\ref{sectconcl}.

% ----------------Section 2 ----------------------

\section{HD~93129A and its surroundings}\label{secttarget}

\subsection{HD~93129A}

HD~93129A is a massive stellar system, comprising at least two stars, with potential spectral types O2\,If*/WN 5 and an O3\,III(f*). The still undefined orbit is wide with a separation of tens of AU and an orbital period of 120 years \citep{gruner2019}.  The coordinates of HD~93129A are R.A.\ (J2000) = $10^{\rm h}~43^{\rm m}~57.458^{\rm s}$ and DEC (J2000)= $-59^{\circ}32\arcmin51.395\arcsec$ and
it is the brightest member of the Trumpler~14 stellar cluster, for which \citet{hur2012} derived a distance of 2.9~kpc. However, the distance of HD~93129A, estimated using the {\it Gaia}~DR3 parallax, is $d = 2.48\pm0.11$~kpc; we will use this last hereafter.
The system had been detected as a radio source decades ago (using the ATCA, with a few MHz bandwidth) despite its distance, as a point source over the course of the time lapse 2003--2009.
%, projects C678, C1726) 
The emission at different bands (1.4 to 24.5~GHz) consisted in flux densities up to 10~mJy, and exhibited a NT spectral index \citep{Benaglia2004,Benaglia2005,Benaglia2006,Benaglia2010}. The data, uniformly processed, demonstrate an increase of the flux density between the epochs 2003--4 and 2008--9, both being well fitted by a power-law spectrum with a steep spectral index (B15). The emission was interpreted as coming mainly from the CWR.

The CWR was identified and mapped through very long baseline interferometry (VLBI) observations conducted with the Australian Long Baseline Array, as documented in B15.  
%(project V191B) 
Re-derived archive {\it HST}/FGS relative positions and VLT data, as cited in \citet{sana2014}, revealed that the separation between two components of HD~93129A was showing a decreasing trend. The data enabled the derivation of a very preliminary periastron passage time towards 2024 (B15). 
Subsequent to a comprehensive spectroscopic investigation, \citet{maiz2017} postulated that the periastron would occur prior to this, in 2017/2018. 
The latest results from optical observations of the system favoured a periastron passage in 2018.70$^{+0.22}_{-0.12}$ \citep{delPalacio2020}.

\subsection{Other sources in the field of view}
 
The Galactic Wolf Rayet Catalogue\footnote{\url{https://pacrowther.staff.shef.ac.uk/WRcat/index.php}} (2024 edition) and the PACWB Catalogue list additional massive stars in systems within the observed %2100
field of view (FoV) of the radio data collected in this work. These are listed in Table 1. Among them is $\eta$\,Car, the brightest system of the Trumpler 16 cluster. 

HD~93129A is situated in the western portion of the Carina Nebula Complex, an extensive region where massive star formation occurs. 
It is characterised by a multitude of radio-emitting matter, organised into a diverse array of structures, spanning a vast range of sizes. In their analysis of deep and detailed radio continuum observations of the entire complex, taken with ATCA in the range of 1--3 GHz with an angular resolution of $\approx 16^{\prime\prime}$, \citet{rebolledo2021} employed a comprehensive approach to describe the region. 

\begin{figure}
   \centering
    \includegraphics[width=9.2cm,angle=0]{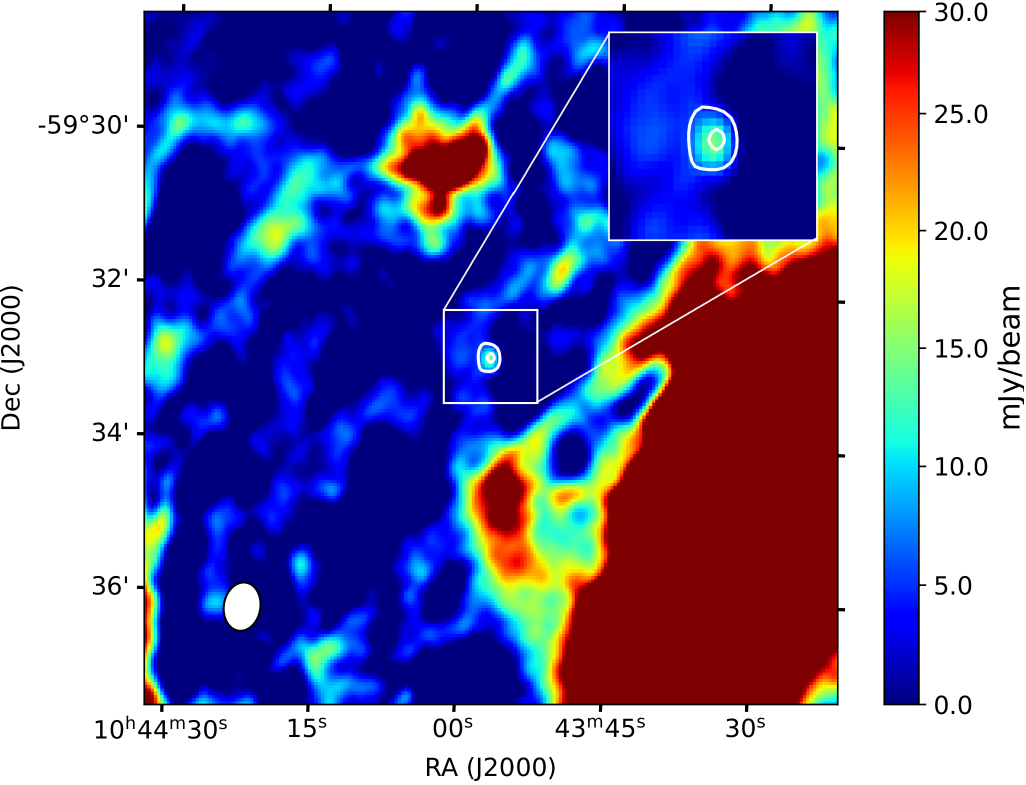}
    \includegraphics[width=9.2cm,angle=0]{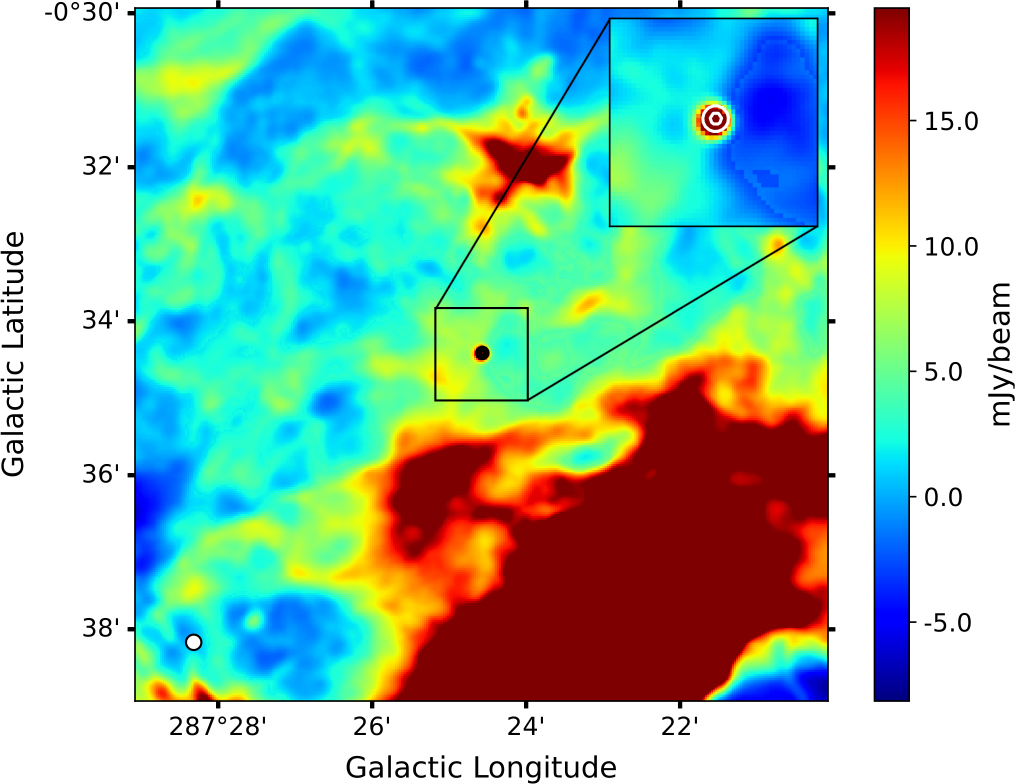}    
      \caption{Continuum emission centred at the position of HD~93129A.
       {\sl Top panel:} at 0.89~GHz {(RACS)}; contour levels over the source of 5.5, %9
      and 13~mJy~beam$^{-1}$; synthesised beam of $15.18'' \times 11.40''$. % PA: $-9.1^\circ$. 
      {\sl Bottom panel}: at 1.36~GHz {(SMGPS)}; contour levels over the source of 20 and 34~mJy~beam$^{-1}$; synthesised beam of $8.0'' \times 8.0''$. {In each panel, the synthesised beam is shown in the bottom left corner, and a zoom of the source is displayed at the top right corner.}} % PA: $8.0^\circ$}
         \label{racslow-mkat}
   \end{figure}

Images of the field of interest were provided from two southern-sky surveys recently published.
One such survey is the Rapid ASKAP Continuum Survey {\citep[RACS,][]{RAC1S2021}}, for which a low-band image centred at 0.89~GHz is publicly available. This image was produced using data obtained on 6 May 2019, which almost coincided with the first epoch of the monitoring reported here. 
The second survey is The SARAO MeerKAT 1.3 GHz Galactic Plane Survey {\citep[SMGPS,][]{MeerKAT2024}}. The data corresponding to the field of HD~93129A were collated on 26 August 2018. MeerKAT survey image cubes contain the stokes-I emission, and also a spectral index map, generated along the bandwidth of 750~MHz.  
Figure~\ref{racslow-mkat} shows the continuum emission of the stellar system  at 0.89~GHz, and at 1.36~GHz, built by the RACS-low and MeerKAT teams, respectively.

%--------------------------Table 1------------
\begin{table}[t]
\caption{Massive binary systems in the field of HD~93129A.} 
\label{tab-massivestars}      
\centering          
\begin{tabular}{l c c c}     % 7 columns
\hline\hline 
Star/system & Spectral class. & Distance & Cluster \\
  &  & (kpc) & or Assoc.\\
\hline                    
   WR~22$^a$ & WN5–7h + O & $2.4\pm{0.2}$ & Car~OB1 \\  
   WR~25$^{a,b}$ & Of/WN + O  & $2.3\pm{0.1}$ & Tr 16   \\
   HD~93250$^{c,d}$ & O4III + O4III  & $2.43\pm0.12$ & Tr 16 \\ 
   $\eta$\,Car$^{e,f}$ & LBV-like object + ? & $2.35\pm0.05$ & Tr 16 \\ 
%   WR~28$^a$  & WN6–8 + O & $6.9\pm0.5$ & \\ % MS 2
\hline                  
\end{tabular}
\tablefoot{References. $a$: \citet{sander2012}; $b$: \citet{hainich2014}; $c$: \citet{Lebouquin2017}; $d$: from Gaia-DR3; $e$: \citet{damineli2008}; $f$: \citet{smith2006}. See also \citet{wrcatalog2023}.}
\end{table}
%______________________________________________

%------------------------ Figure 1 ---
%______________________________________________

% -----------------Section 3 -------------------

\section{Observations, calibration and images}\label{sectobs}

\begin{figure}
   \centering
   \includegraphics[width=7cm]{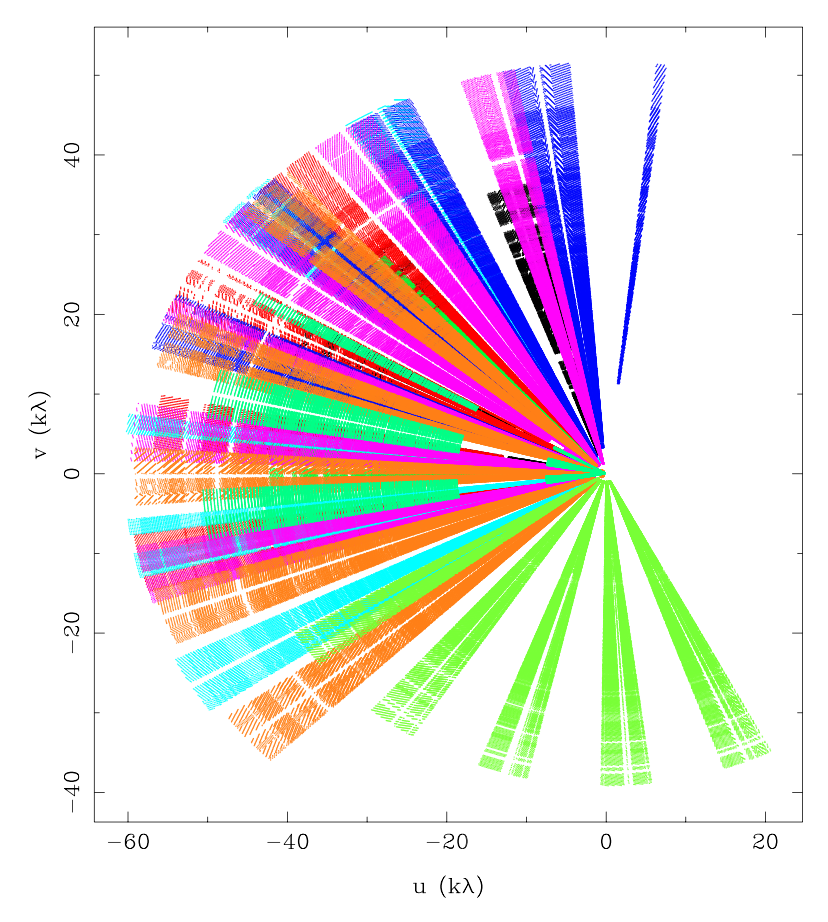}
      \caption{Coverage of the $uv$-plane at 2.1~GHz, including data from all epochs, from points coloured by epoch. One out of every 100 points has been plotted.}
         \label{Fig-2100uvcoverage}
   \end{figure}

The data were collected with the Australia Telescope Compact Array over a period of 17 months, from May 2019 to September 2020. They were gathered on 11 dates/epochs in two distinct ATCA bands: the so-called `16~cm' and `4~cm' bands, which encompass classical bands S, C and X, and were obtained in continuum mode.  The S, C, and X bands were centred at 2.1, 5.5, and 9.0 GHz, with bandwidths of 2~GHz. The half-power beam widths at the centre of the observation bands are $44.5^\prime$, $18^\prime$ and $12^\prime$, respectively. The correlator configuration was CFB~1M-0.5k (2048 channels). With the exception of the initial epoch, the observations were carried out remotely. At both the C- and X-band frequencies, the observations were conducted in a simultaneous manner. Table~\ref{observingparm} presents the essential details regarding the observations, including the date, the centre of the observed bands, the array configuration, the duration of each observation run, and the time on the source HD 93129A. The 11 epochs spanned 500 days, with an average separation of 50 days.
Figure~\ref{Fig-2100uvcoverage} illustrates the $uv$ coverage for S-band across all observation epochs; similar results were obtained for C and X bands.

%-------------------------- Figure 2 -------
%______________________________________________

In the initial phase of the observations, the ATCA setup was configured to either utilise the well-established and prominent source 1934$-$638 or the source 0823$-$500, if the former was not sufficiently elevated above the horizon. The source 1934$-$638 was observed as the primary calibrator (this includes bandpass). The source PKS~1059$-$63, which has been used extensively by our research group in previous campaigns on the same target, served as the gain and phases calibrator.

The data were calibrated in accordance with standard procedures\footnote{See for example \url{https://www.narrabri.atnf.csiro.au/observing/users_guide/html/atug.html}} using the {\sc miriad} package \citep{sault1995}. The images of the phase calibrator were subjected to a consistency analysis, and further flagging on all data for certain days conducted using the \texttt{pgflag} commands. Each epoch's data set was calibrated individually. Visualisation and analysis were facilitated by the {\sl karma} software, as referenced in \citet{gooch1996}.

\subsection{Imaging of HD~93129A}

The complete dataset from August 2019, the sole instance of the 750* array configuration, was unable to be properly imaged. This was primarily due to the presence of considerable RFI, in addition to the fact that the baseline for this specific configuration did not provide sufficient \textit{uv} coverage to overcome the challenge of imaging the extensive structures present within the FoV.  
Similarly, the September 2019 dataset in the 9.0~GHz band was also excluded due to the presence of significant RFI. Furthermore, the S-band was not observed in April and September 2020. For a comprehensive overview of the observational parameters, we refer to Table~\ref{observingparm}.

\begin{figure}
   \centering
   \includegraphics[width=7cm]{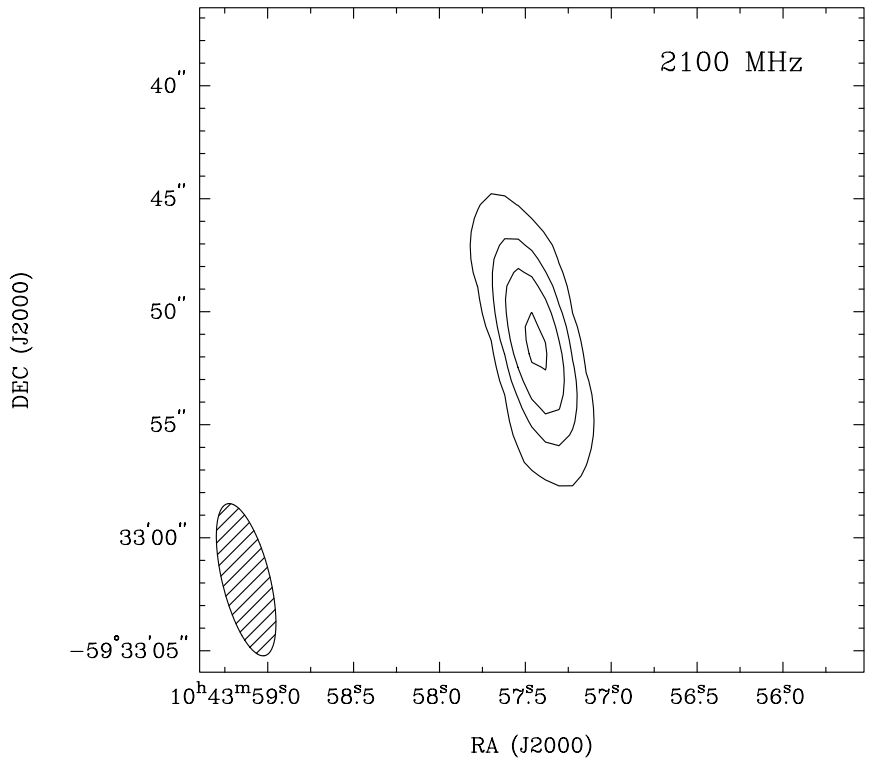}
     \includegraphics[width=7cm]{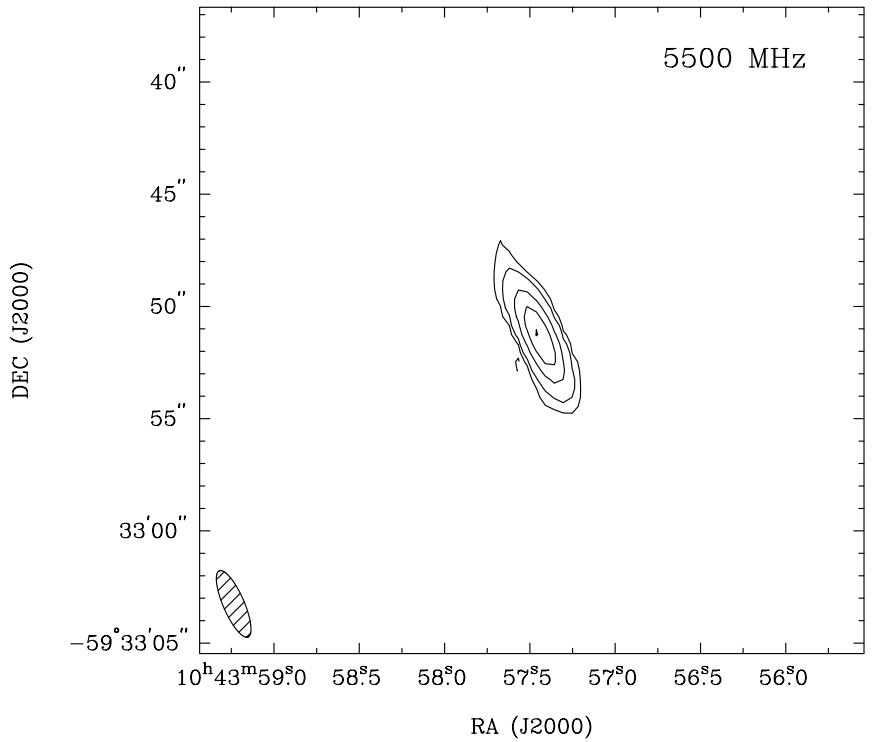}
       \includegraphics[width=7cm]{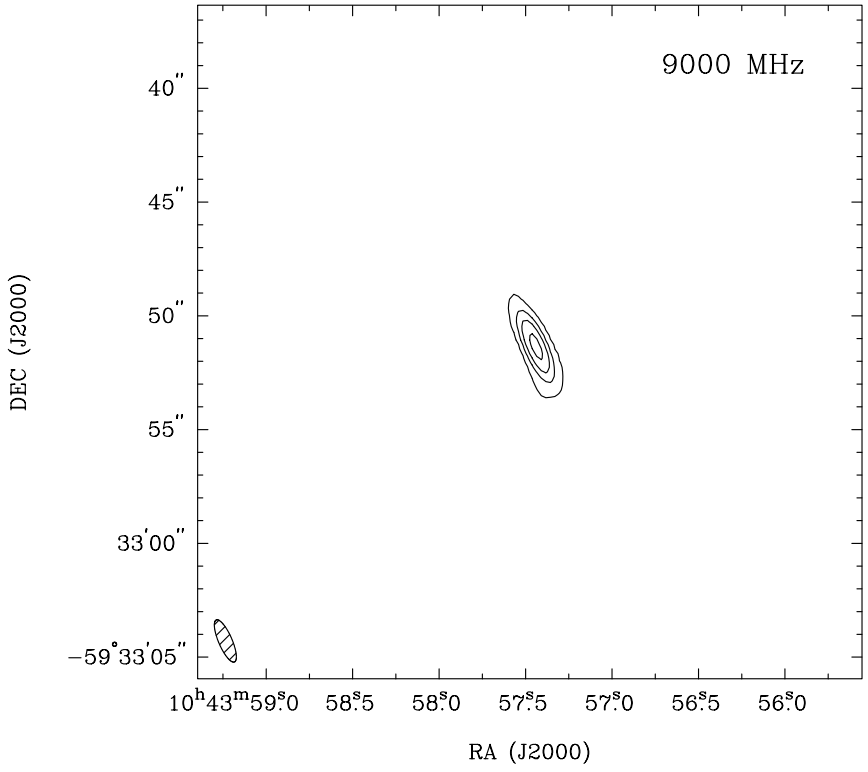}
      \caption{Continuum emission at the position of HD~93129A, observed in June 2019. {\sl Top panel}: at 2.1~GHz; contour levels: 3, 10, 20 and 35, in units of $\sigma$ (=0.55~mJy~beam$^{-1}$). {\sl Central panel}: at 5.5~GHz; contour levels: 3, 10, 40, 90 and 150, in units of $\sigma$ (=0.12~mJy~beam$^{-1}$). {\sl Bottom panel}: at 9.0~GHz; contour levels:  3, 20, 50 and 110, in units of $\sigma$ (=0.10~mJy~beam$^{-1}$). The corresponding synthesised beams are shown in the bottom left corner.}
         \label{3panels}
   \end{figure}

The images centred on HD~93129A were created using the software package {\sc miriad}. The combination of robust weighting and a limited $uv$-range (0 and $\geq$ 3~k$\lambda$) yielded the optimal images. Due to differences in array configurations, observing times, and differential radio interference, distinct strategies were employed for each epoch across all bands. Figure~\ref{3panels} illustrates the images obtained for HD~9319A at the three observed bands corresponding to the June 2019 epoch.

%-------------------------- Figure 3 ---
%______________________________________________

\subsection{Imaging of the HD~93129A field}\label{subsecImages}

Images illustrating the complete field of view, by using the full set of data, were generated with the Common Astronomy Software Applications \citep[CASA,][]{CASA2007}, at the three observed bands. The observations were conducted using a very wide spectral bandwidth ($\sim$2~GHz): the results demonstrated that the optimal outcome was consistently achieved through the utilisation of the Multi-Scale, Multi-Term Frequency Synthesis option of the CASA task \texttt{tclean}, in conjunction with Briggs weighting of $+0.5$, and discarding larger wavelengths. 

\begin{figure*}
\centering
\includegraphics[width=18cm]{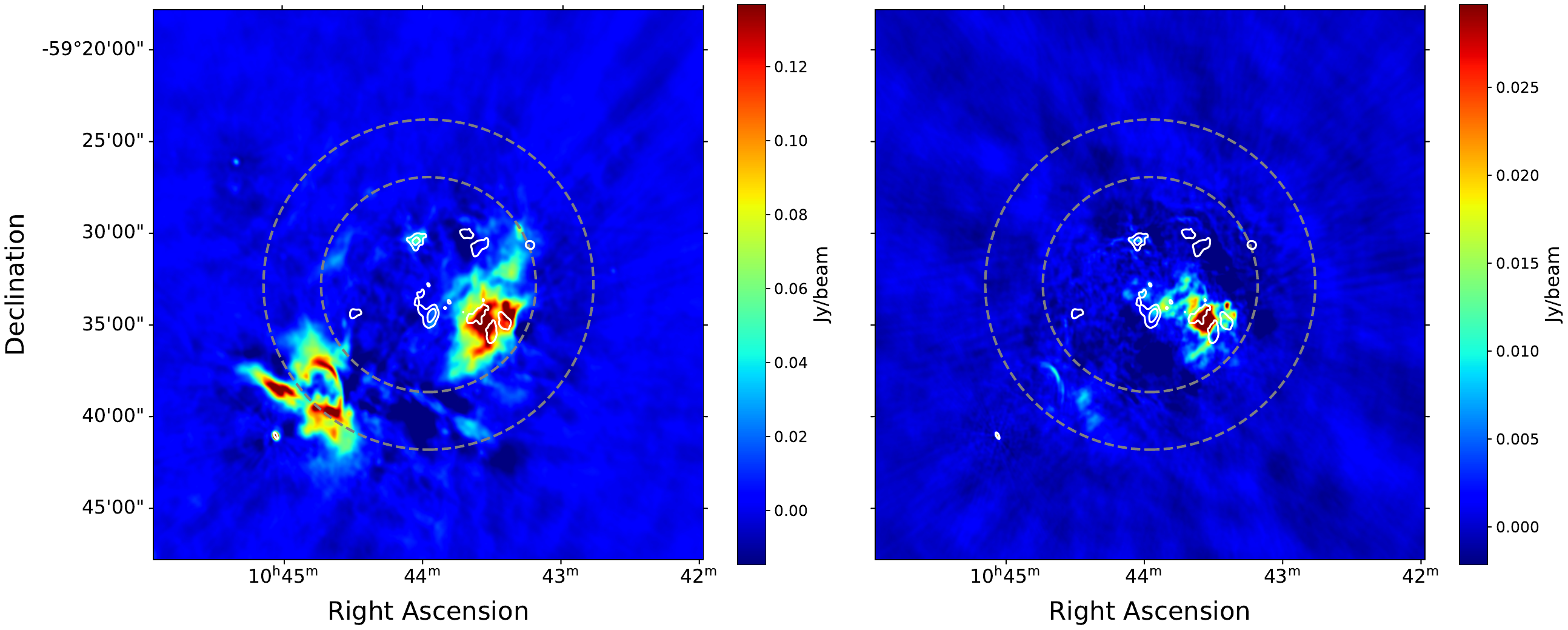}
\caption{Continuum emission at 2.1~GHz ({\sl left panel}) and 5.5~GHz ({\sl right panel}), in colour scale, without correction for primary beam. In both panels, the HPBW at 5.5~GHz (larger) and 9.0~GHz (smaller) are marked with dashed lines, while the HPBW at 2.1~GHz is outside the images. The emission at 9.0~GHz is shown in white contours, with levels of 6 (3$\sigma$), 23.6 and 41.3 mJy~beam$^{-1}$. The small and closed contour level at the centre of the images represents the location of HD~93129A. The almost circular source towards the bottom left corner is $\eta$\,Car.}
\label{fig:atcafovimages}
\end{figure*}

The synthesised beams of the final continuum emission images resulted in a resolution of $16.3''\times 12.3''$ at 2.1~GHz, $10.7''\times 7.7''$ at 5.5~GHz and $1.1''\times 0.7''$ at 9.0~GHz. The images are displayed in  Fig.~\ref{fig:atcafovimages}.

A complementary set of images was produced in the three bands but retaining the longest wavelengths, such as \textit{uv}-range $\geq$~40\,k$\lambda$ for the image at 2.1~GHz and \textit{uv}-range $\geq$ 30\,k$\lambda$ for the other two. The aim was to optimise the detection of discrete sources when looking for emission from other CWBs in the field (see Table~\ref{tab-massivestars}).

%------------------Figure 4 ------------
%______________________________________________

%---------------------------- Table 2 ------------
%
\begin{table*}
\caption{Main parameters of the observations.}            
\label{observingparm}      
\centering          
\begin{tabular}{c c c c c c c c c c}     % 11 columns
\hline\hline 
Observing & Array  & Total time  & \multicolumn{3}{c}{Central frequency } && \multicolumn{3}{c}{$t_{\rm os}$}\\
date &  configuration & (h) &\multicolumn{3}{c}{(GHz)} &&  \multicolumn{3}{c}{(h)}\\
\hline                    
   10-05-2019 & 1.5 B & 5.81 & 2.1 & 5.5 & 9.0 && 1.97 & 1.97 & 1.97 \\  
   21-06-2019 & 6 A   & 5.64 & 2.1 & 5.5 & 9.0 && 1.97 & 1.97 & 1.97 \\
   19-08-2019 & 750 C & 3.77 & 2.1 & 5.5 & 9.0 && 1.13 & 0.99 & 0.99 \\%data-lost
   16-09-2019 &  6 C  & 6.20 & 2.1 & 5.5 & 9.0 && 2.39 & 1.81 & 1.81 \\
   27-11-2019 & 1.5 C & 3.58 & 2.1 & 5.5 & 9.0 && 1.24 & 0.99 & 0.99 \\
   10-01-2020 & 6 A   & 5.26 & 2.1 & 5.5 & 9.0 && 2.06 & 1.89 & 1.89 \\
   12-02-2020 & 6 A   & 6.19 & 2.1 & 5.5 & 9.0 && 2.61 & 1.97 & 1.97 \\
   12-03-2020 & 6 D   & 6.31 & 2.1 & 5.5 & 9.0 && 2.46 & 2.22 & 2.22 \\
   09-04-2020 & 6 A   & 1.62 & --  & 5.5 & 9.0 &&  --  & 1.23 & 1.23 \\
%  14-05-2020 & --  & 5.5 & 9.0  & 567& 55 &\\
   10-06-2020 & 1.5 C & 5.89 & 2.1 & 5.5 & 9.0 && 2.39 & 1.97 & 1.97 \\
   24-09-2020 &  6 B  & 1.49 & --  & 5.5 & 9.0 && -- & 0.97 & 0.97 \\
\hline                  
\end{tabular}
\tablefoot{The Total time refers to the amount of telescope time dedicated to all sources, including calibrators, across all observed bands. The time on target, represented by $t_{\rm os}$, pertains to the duration spent observing HD~93129A at each observed band.}
\end{table*}
% Total time: taken from uvindex log
%__________________________

% -------------------Section 3-------------------

\section{Results}\label{sectresults}

\subsection{The light curve of HD~93129A}

\begin{figure}
\centering
\includegraphics[width=\linewidth]{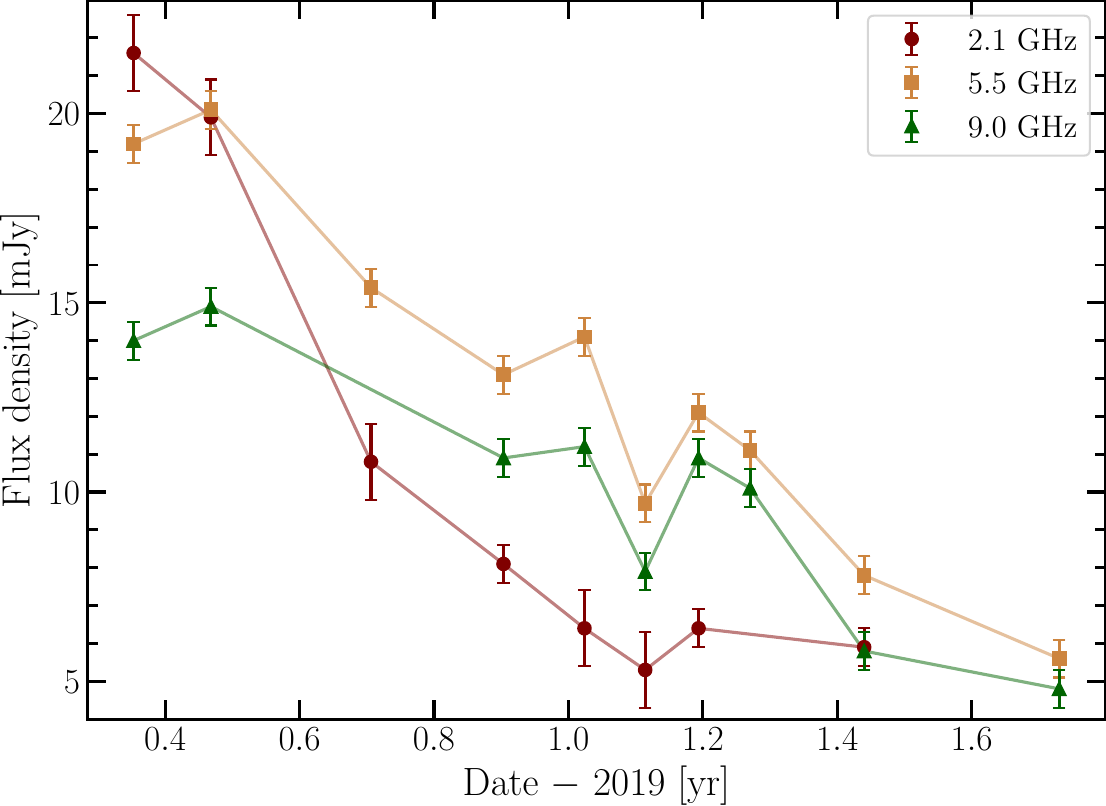} 
\includegraphics[width=\linewidth]{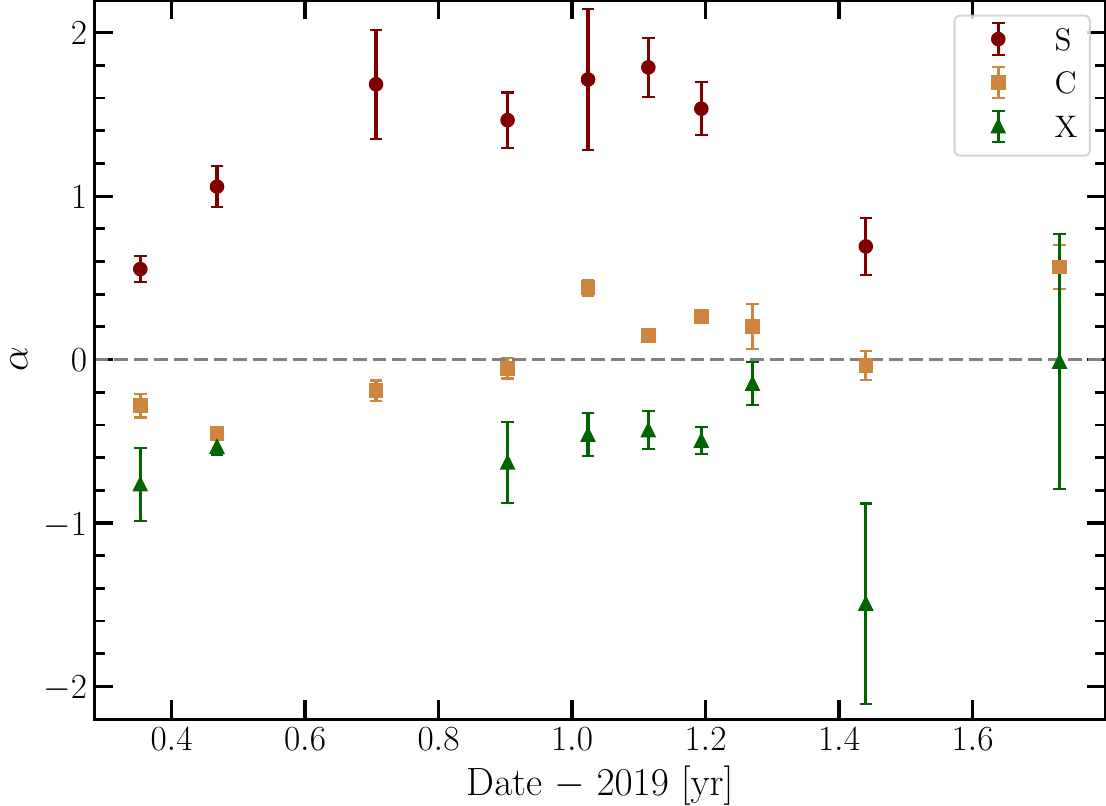} 
\caption{Light curves of HD~93129A representing the flux densities ({\sl top panel}) and  evolution of the spectral index ({\sl bottom panel}), as a function of time, for the three ATCA bands observed in this work. {The $x$ axis values correspond to dates starting from 2019.0}.}
\label{fig:lightcurves}
\end{figure}

Table~\ref{subbandfluxes} (third column), presents the flux density values of HD~93129A, measured at the three observed bands over the course of the monitoring period. A Gaussian fit was carried out on the flux density of the radio source at the position of HD~93129A with the aid of CASA. The fits were found to correspond to discrete sources. The root-mean-square ($rms$) was measured in a ring surrounding the radio source. In all cases, the integrated flux density and the peak flux density of the fit yielded comparable values. The mean of these values, along with the difference, is presented together with the image $rms$ error. The corresponding light curves are shown in Fig.~\ref{fig:lightcurves} (top panel).

In addition, taking advantage of the Compact Array Broadband Backend (CABB), we obtained flux densities of the stellar system in four subbandds (here named SB1, SB2, SB3 and SB4) at the three observed bands (S, C and X), for most of the epochs; they are listed also in Table~\ref{subbandfluxes}. With these data we calculated the average spectral index $\alpha$ (defined as $S_\nu \propto \nu^\alpha$) at each band and at each epoch; the spectral index evolution is shown in Fig.~\ref{fig:lightcurves} (bottom panel).

The flux density of HD~93129A measured in the RACS-low image (0.89~GHz), as observed in May 2019, was found to be $10.4\pm1.5$~mJy; the synthesised beam corresponding to these coordinates was $15.18'' \times 11.40''$. %PA=-9.1 deg
Regarding the MeerKAT image (1.36~GHz), the synthesised beam is $8'' \times 8''$, the flux density at the position of  HD~93129A is $40\pm3$~mJy, and the spectral index {value of the peak} is $\alpha = -0.69$.
%\textcolor{red}{\st{However, it is likely that the spectral energy distribution (SED) is flatter at 1.3~GHz due to absorption effects, suggesting a more negative spectral index towards S band. To account for this, we adopted a value of $\alpha = -0.9 \pm 0.2$. We then derived a flux at 2.1~GHz of $S_{2.1} = 27\pm3$~mJy, for the Aug 2018 observation date.}}

%---------------------------- Table 3 ------------
\begin{table*}
\caption{Flux density values of HD~93129A at full bands and subbands, along the epochs.}     
\label{subbandfluxes}      
\centering          
\begin{tabular}{l c c c c c c c c c c}  % 7 columns
\hline\hline 
Observing  & Band &  $S_{\rm full-band}$ & $\nu_{\rm SB1}$ & $S_{\nu,{\rm SB1}}$ &  $\nu_{\rm SB2}$ & $S_{\nu,{\rm SB2}}$ &  $\nu_{\rm SB3}$ & $S_{\nu,{\rm SB3}}$ &  $\nu_{\rm SB4}$ & $S_{\nu,{\rm SB4}}$\\
 date &  & (mJy)  &   (GHz) & (mJy) &   (GHz) & (mJy) &   (GHz) & (mJy) & (GHz) & (mJy) \\
\hline                    
10-05-2019 & S & 21.6$\pm$1.0 & 1.49 & 16.8$\pm$0.5 & 1.94 & 20.3$\pm$1.0 & 2.38 & 23.0$\pm$1.0 & 2.85 & 23.7$\pm$1.0\\  
           & C & 19.2$\pm$0.5 & 4.81 & 19.6$\pm$1.0 & 5.27 & 19.6$\pm$0.5 & 5.73 & 18.9$\pm$0.5 & 6.19 & 18.4$\pm$0.5\\  
           & X & 14.0$\pm$0.5 & 8.31 & 14.8$\pm$1.2 & 8.77 & 14.5$\pm$1.0 & 9.20 & 13.4$\pm$0.5 & 9.69 & 13.3$\pm$0.5\\  
21-06-2019 & S & 19.9$\pm$1.0 & 1.49 & 13.1$\pm$0.5 & 1.94 & 17.6$\pm$0.5 & 2.38 & 23.3$\pm$0.5 & 2.85 & 25.2$\pm$1.0\\  
           & C & 20.1$\pm$0.5 & 4.81 & 21.0$\pm$0.5 & 5.27 & 20.1$\pm$0.5 & 5.73 & 19.3$\pm$0.5 & 6.19 & 18.8$\pm$1.0\\  
           & X & 14.9$\pm$0.5 & 8.31 & 15.3$\pm$0.5 & 8.77 & 15.0$\pm${0.5} & 9.20 & 14.6$\pm$0.5 & 9.69 & 14.1$\pm$0.5\\   
16-09-2019 & S & 10.8$\pm$1.0 & 1.49 &  4.4$\pm$0.5 & 1.94 &  8.9$\pm$0.5 & 2.38 & 12.1$\pm$0.5 & 2.85 & 13.8$\pm$1.0\\
           & C & 15.4$\pm$0.5 & 4.81 & 15.6$\pm$0.5 & 5.27 & 15.0$\pm$0.5 & 5.73 & 15.0$\pm$0.5 & 6.19 & 14.8$\pm$0.5\\
           & X & ** &&&&&&&&\\  %RFI
27-11-2019 & S & 8.1$\pm$0.5 & 1.49 &  3.9$\pm$0.5 & 1.94 &  5.2$\pm$0.5 & 2.38 &      **      & 2.85 & 9.8$\pm$1.0\\
           & C & 13.1$\pm$0.5 & 4.81 & 12.6$\pm$0.5 & 5.27 & 12.8$\pm$0.5 & 5.73 & 12.5$\pm$0.5 & 6.19 & 12.5$\pm$0.5\\
           & X & 10.9$\pm$0.5 & 8.31 & 11.5$\pm$0.5 & 8.77 & 11.2$\pm$0.5 & 9.20 & 10.4$\pm$0.5 & 9.69 & 10.6$\pm$0.5\\  
10-01-2020 & S &  6.4$\pm$1.0 & 1.49 &  2.6$\pm$0.5 & 1.94 &  5.2$\pm$0.5 & 2.38 &  8.5$\pm$0.5 & 2.85 &  9.0$\pm$0.5\\
           & C & 14.1$\pm$0.5 & 4.81 & 12.8$\pm$0.5 & 5.27 & 13.5$\pm$0.5 & 5.73 & 14.0$\pm$0.5 & 6.19 & 14.3$\pm$0.5\\
           & X & 11.2$\pm$0.5 & 8.31 & 12.1$\pm$0.5 & 8.77 & 11.6$\pm$0.5 & 9.20 & 11.3$\pm$0.5 & 9.69 & 11.3$\pm$0.5\\ 
12-02-2020 & S &  5.3$\pm$1.0 & 1.49 &    **        & 1.94 &  3.5$\pm$0.5 & 2.38 &  5.4$\pm$0.5 & 2.85 &  7.1$\pm$0.5\\
           & C &  9.7$\pm$0.5 & 4.81 &  8.7$\pm$0.5 & 5.27 &  8.7$\pm$0.5 & 5.73 &  8.9$\pm$0.5 & 6.19 &  9.0$\pm$0.5\\
           & X &  7.9$\pm$0.5 & 8.31 &  8.1$\pm$0.5 & 8.77 &  7.8$\pm$0.5 & 9.20 &  7.6$\pm$0.5 & 9.69 &  7.6$\pm$0.5\\  
12-03-2020 & S &  6.4$\pm$0.5 & 1.49 &  2.5$\pm$0.5 & 1.94 &  4.5$\pm$0.5 & 2.38 &  5.8$\pm$0.5 & 2.85 &  7.4$\pm$0.5\\
           & C & 12.1$\pm$0.5 & 4.81 & 11.3$\pm$0.5 & 5.27 & 11.7$\pm$0.5 & 5.73 & 11.9$\pm$0.5 & 6.19 & 12.1$\pm$0.5\\
           & X & 10.9$\pm$0.5 & 8.31 & 11.2$\pm$0.5 & 8.77 & 10.8$\pm$0.5 & 9.20 & 10.5$\pm$0.5 & 9.69 & 10.4$\pm$0.5\\ 
09-04-2020 & S &  -- & &&&&&&&\\ % not observed
           & C & 11.1$\pm$0.5 & 4.81 & 10.4$\pm$0.5 & 5.27 & 11.1$\pm$0.5 & 5.73 & 10.8$\pm$0.5 & 6.19 & 11.1$\pm$0.5\\
           & X & 10.1$\pm$0.5 & 8.31 & 10.4$\pm$0.5 & 8.77 & 10.6$\pm$0.5 & 9.20 & 10.4$\pm$0.5 & 9.69 & 10.2$\pm$0.5\\ 
10-06-2020 & S &  5.9$\pm$0.5 & 1.49 &    **        & 1.94 &  4.7$\pm$0.5 & 2.38 &  5.1$\pm$0.5 & 2.85 &  6.1$\pm$0.5\\
           & C &  7.8$\pm$0.5 & 4.81 &  7.3$\pm$0.5 & 5.27 &  7.3$\pm$0.5 & 5.73 &  7.1$\pm$0.5 & 6.19 &  7.3$\pm$0.5\\  
           & X &  5.8$\pm$0.5 & 8.31 &  6.0$\pm$0.5 & 8.77 &  5.6$\pm$0.5 & 9.20 &  5.5$\pm$0.2 & 9.69 &  4.5$\pm$0.5\\
24-09-2020 & S & -- & &&&&&&&\\ % not observed
           & C & 5.6$\pm$0.5 & 4.81 &  5.0$\pm$0.5 & 5.27 &  5.5$\pm$0.5 & 5.73 &  5.7$\pm$0.5 & 6.19 & 5.8$\pm$0.5\\
           & X &  4.8$\pm$0.5 & 8.31 &  4.6$\pm$0.5 & 8.77 &  3.9$\pm$0.5 & 9.20 &  4.5$\pm$0.5 & 9.69 &  4.3$\pm$0.5\\ 
\hline                  
\end{tabular}
\tablefoot{SB1--SB4 denote the first to fourth subbands, ordered by increasing frequency.
--: not observed. **: ill image or data with severe RFI.}
\end{table*}
%__________________________
% Computed from miriad images with CA06, and using CASA
% values taken from file Flux-dens-allsources-last.numbers
% figure taken from file Flux-dens-hd.numbers
%

%--------------------- Figure 5 ------------
%__________________________

\subsection{The field of view of HD~93129A}

Of the three observed ATCA bands, the one centred at 2.1~GHz, having the largest field of view, was the only that fully encompassed the source $\eta$\,Car and its surroundings. Using the image provided by RACS-low, centred at 0.89~GHz, we computed the spectral index distribution and the corresponding error maps between the two images. To this end, we convolved and regridded both images to common angular resolution of $20'' \times 20''$ and pixel and image sizes. 

\begin{figure}
   \centering
   \includegraphics[width=9.3cm]{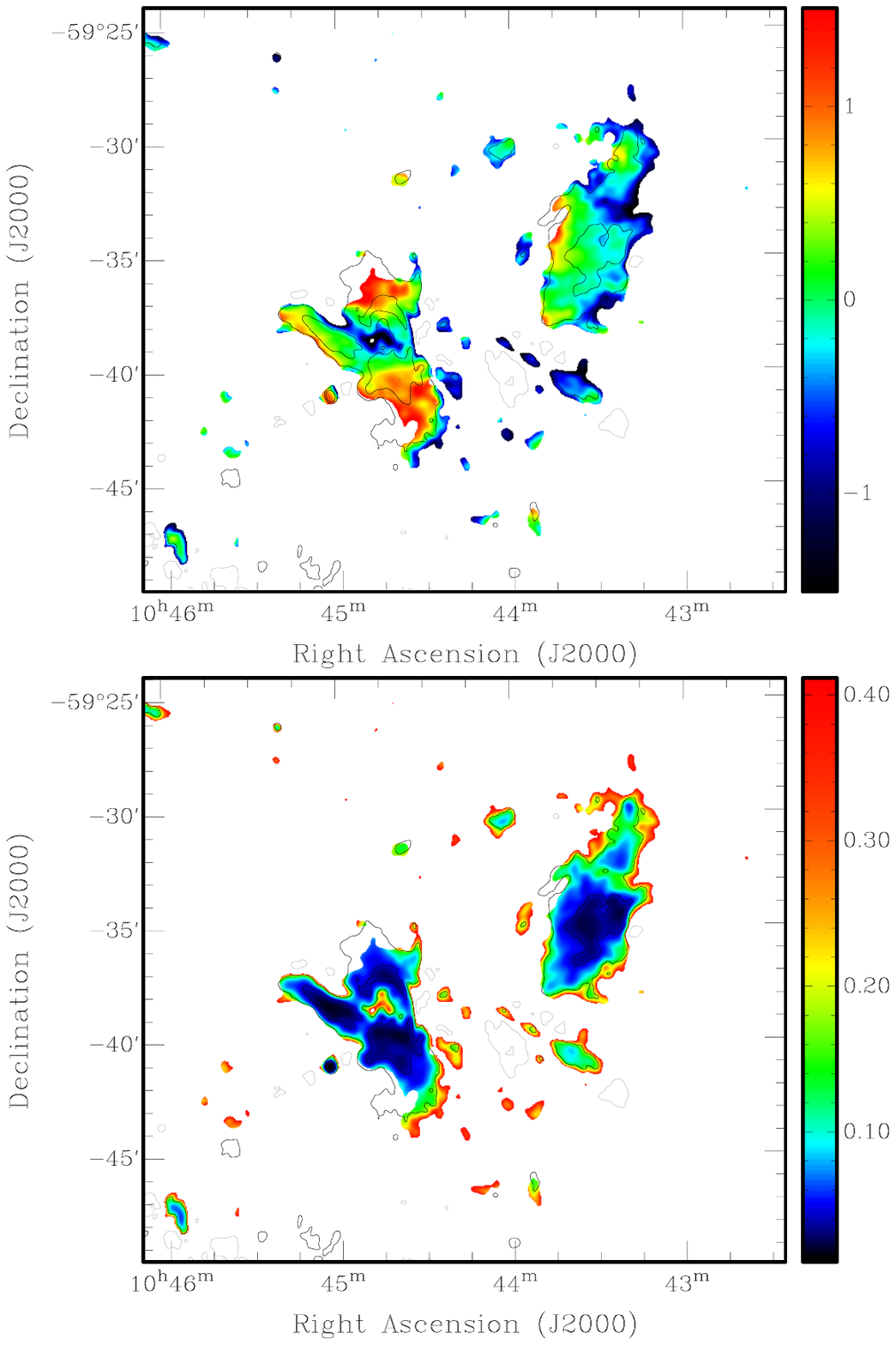}
      \caption{Distribution of the spectral index ({\sl top panel}) and its error ({\sl bottom panel}), between data at 0.89~GHz (RACS-low) and 2.1~GHz (ATCA). The contour lines represent the 2.1~GHz emission levels at $-$25 (in grey colour), 25 and 100~mJy~beam$^{-1}$ (in black colour).}
         \label{spixanderror}
   \end{figure}

To calculate the spectral index map and error distribution, we considered the RACS-low and 2.1-GHz ATCA image pixels above 5$\sigma$; $\sigma$ was taken as the $rms$ of the image, with values 4 and 5~mJy~beam$^{-1}$ for ATCA and RACS images, respectively. The resulting spectral index distribution and its associated error map are presented in Fig.~\ref{spixanderror}.

%----------------------------- Figure 6 ---
%__________________________

%------------------------------------ Section 5

\section{Discussion on the system HD~93129A}\label{sectdisc} 

\subsection{Indication of periastron from radio observations}\label{sec-perifromradio}

\citet{maiz2017} presented precise astrometric measurements of the system HD~93129A taken with different instruments from 1996 to 2016. The analysis of the data allowed to predict a periastron passage for 2017/2018. The results also suggested that the primary star itself could be a tight pair.
As part of the study of \citet{delPalacio2020}, the authors quote a periastron passage derived with data up to 2017 of 2018.54$^{+0.54}_{-0.32}$, but they also considered supplementary (consecutive) data to that of \citet{maiz2017}, at some epochs between 2017--2019. With this larger database, \textcolor{red}{\citet{delPalacio2020}} updated the proposed periastron time to 2018.70$^{+0.22}_{-0.12}$, and also constrained the separation of the components at periastron to $7.91\pm0.42$~mas (or 19.6$\pm$1.0~AU for the distance adopted here).

As a first approximation, the intrinsic synchrotron radio emission from the CWR is expected to increase as periastron approaches, but the enhanced FFA from the stellar wind material can significantly compete with this intrinsic increase. In extreme situations it may even be suppressed, depending on orbit inclination and stellar wind properties \citep[see e.g.][]{DeBecker2019}. Away from periastron, the synchrotron emission region gradually emerges from the thick wind and may be measured, even though it intrinsically drops with increasing separation.

\begin{figure}
\centering
\includegraphics[width=8.5cm]{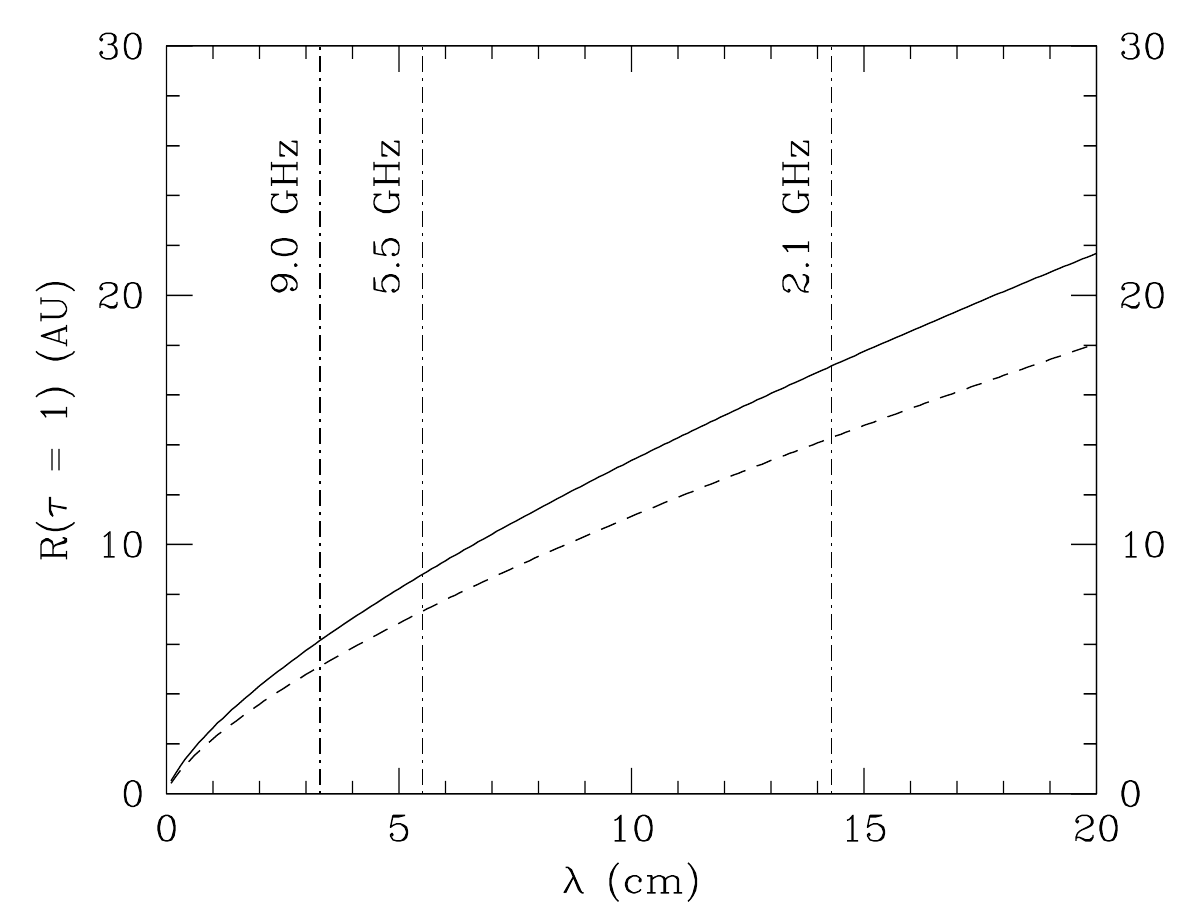}
\caption{Photosphere radius of the two main components of HD~93129A system as a function of the observing wavelength; solid line: HD~93129Aa; dashed line: HD~93129Ab. The wavelengths corresponding to the observed ATCA bands are shown with dash-dotted lines.}
\label{fotosferas}
\end{figure}

The radio photosphere of a star can be calculated on the basis of the formalism presented by \citet{wrightbarlow1975}, considering an unit optical depth of a smooth wind as used in several previous studies \citep[e.g.][]{Benaglia2019,DeBecker2019,Blanco2024}. 
In Fig.~\ref{fotosferas} we have plotted the photosphere radius of each binary component of HD~93129A, as a function of wavelength. The wind parameters assumed in the calculations are as follows \citep[e.g.][]{delPalacio2020}: mass-loss rates $\dot{M_{\rm a}} = 10^{-5}$~M$_\odot$~yr$^{-1}$ and $\dot{M_{\rm b}} = 0.6 \times 10^{-5}$~M$_\odot$~yr$^{-1}$, and terminal wind velocities $v_{\infty,{\rm a}} = 3200$~km~s$^{-1}$ and $v_{\infty,{\rm b}} = 2800$~km~s$^{-1}$,\footnote{{Given the very large orbital separations, the winds reach their terminal velocity much before reaching the CWR, and thus the acceleration profile is not important.}}. We also adopted a mean molecular weight $\mu= 1.5$ (common value for O-type and early WN), a mean number of electrons per ion $Z=1.0$ (same), and a $rms$ ionic charge $\gamma = 1.0$ (same). {These numbers are in the range of values adopted for instance by \citet{Leitherer1995} and \citet{Montes2009} for similar massive star winds.} For the effective temperature of the stars, we considered $T_{\rm eff,a} = 50000$~K and $T_{\rm eff,b} = 43000$~K, {and for their radii we adopted $R_{a} = 18.3$~R$_\odot$ and $R_{b} = 16.6$~R$_\odot$,} which constitute valid approximations for their not well-determined classification \citep{Martins2005, Crowther2007, Muijres2012}. {The temperature in the ionised winds is assumed to be $0.3 T_\mathrm{eff}$ \citep[e.g.][]{Drew1990}.} 

Figure~\ref{fotosferas} shows that the 2.1~GHz radio photosphere is of the order of the stellar separation at periastron as quoted in \citet{delPalacio2020}. We caution that the curves in Fig.~\ref{fotosferas} should not be seen as strict values, but as indicative of an order of magnitude. In particular, the radio photosphere size is sensitive to the uncertainty on the mass loss rate ($R \propto {\dot M}^{2/3}$, \citealt{wrightbarlow1975}). What it tells us is that, at the time of our observations, we expect some substantial modulation of the synchrotron emission by FFA. However, a significant part of the emission region is out of the thickest region of the winds, warranting the measurement of synchrotron emission from the system even close to periastron. We also note that, within the uncertainties on the wind parameters, both winds are expected to contribute with similar strength to FFA. One wind can however contribute significantly more depending on the specific orientation of the system. In particular, a severe drop of the synchrotron emission close to periastron is expected if a strongly absorbing wind is in front of the emission region.

At first glance, the radio light curves in Fig.~\ref{fig:lightcurves} seem to indicate that periastron {would have occurred prior to May-June 2019}, and that its effects were still noticeable until the end of the observations presented here (September 2020).
However, this simple picture is limited by the uncertainties on several key parameters, such as the stellar wind parameters, the inclination and eccentricity of the orbit. All these factors need to be well determined to fully understand the radio results of HD~93129A and characterise its behaviour. Below we address these issues more quantitatively by means of detailed modelling.

%------------------ New Figure 7-ex Figure 11
%__________________________

\subsection{SED near periastron} \label{sec:sed_may2019}

We focus first on the epoch of May 2019, for which we have broad spectral coverage; we also note that this epoch is relatively close to the periastron passage epoch proposed in \citet{delPalacio2020} ({ocurring} by the end of 2018).
The expected orbital separation during this epoch is $D \approx 19$\ AU, with a projection angle $\psi \sim 40^\circ$ \citep{delPalacio2020}. To interpret the SED, we use the non-thermal emission model for colliding-wind binaries presented in \cite{delPalacio2020}. {This model considers the CWR as a two-dimensional axisymmetric surface. On each side of the surface there is a thin shock produced by the incoming stellar wind. Relativistic particles are accelerated in the shocks and injected with a power-law distribution of index $p=3$ (with $Q(E) \propto E^{-p}$) and transported along the streamlines. The thermodynamical quantities at the shocks are calculated using analytical prescriptions that depend on the parameters of the incoming stellar wind in the pre-shock region. The emission from the relativistic particles by different processes---with synchrotron emission being the most relevant for this work---is obtained at each location of the CWR and then corrected for FFA along the line of sight. We note that only the ionised stellar winds contribute to the FFA, as the hot and dilute gas in the CWR has a very low opacity \citep[e.g.][]{delPalacio2016}. {We also note that the CWR can produce significant free--free emission at high frequencies ($>$20~GHz) in short-period systems with dense CWRs \citep{pittard2010}. However, this is not the case for HD~93129A, as the separation between the stars is large and the shocks in the CWR are adiabatic. Thus, we neglect the thermal emission from the CWR}.}

\begin{figure}[h]
\centering
\includegraphics[width=\linewidth]{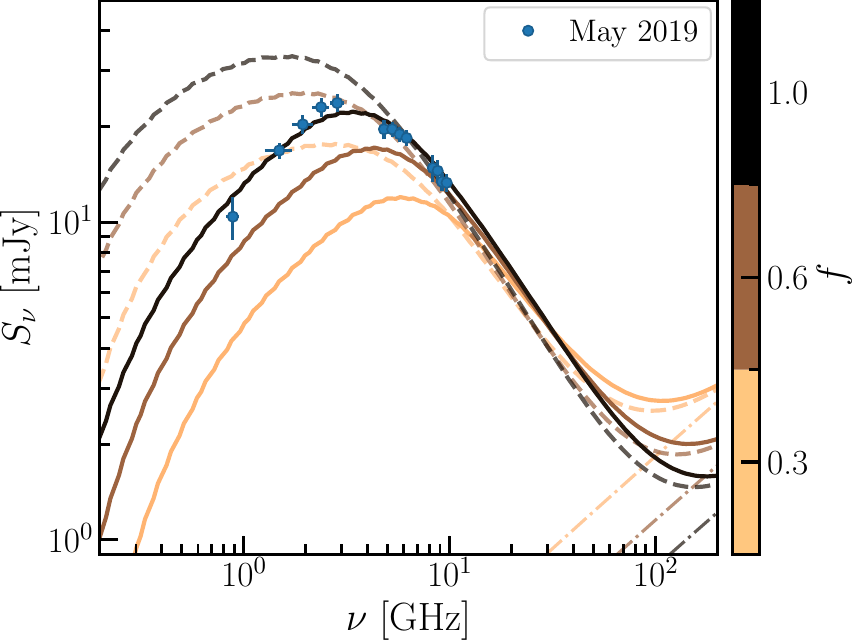}\\
\includegraphics[width=\linewidth]{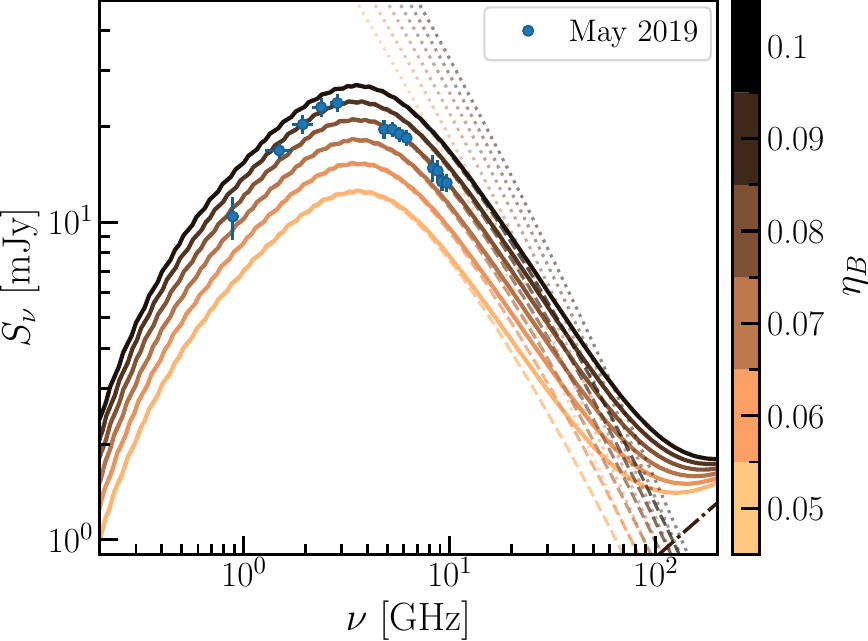}
\caption{Radio SED of HD~93129A observed in May 2019, measured from the RACS-low (May 6th) and  ATCA images (May 10th, this work). The solid lines are the total (absorption-corrected) SEDs, whereas the dashed-dotted lines are the total free--free emission from the combined stellar winds. \textit{Top panel:} Exploration of the impact of the wind volume filling factor $f$ {and the magnetic field prescription}. The dashed lines show the SEDs calculated without including the frozen-in conditions where the shocked fluid becomes transonic.  
\textit{Bottom panel:} Exploration of the impact of the magnetic field strength (defined via the parameter $\eta_B$; Sect.~\ref{sec:sed_may2019}). The dotted lines represent the unabsorbed synchrotron emission.}
\label{fig:sed_may2019}
\end{figure}

{Special attention is given to how the magnetic field intensity is estimated in the CWR. For systems in which the CWR lies at distances much larger than the stellar radii, the dominant component of the stellar magnetic field is the toroidal one, which decays with the distance to the star as $r^{-1}$. This magnetic field is then amplified by adiabatic compression in the shocks. However, additional processes such as magnetic field amplification can also be at play. For this reason, in the original model from \cite{delPalacio2016} a phenomenological approach was taken. The ratio between the magnetic and thermal pressures ($P_\mathrm{mag}$, $P_\mathrm{th}$) is assumed to be constant along the CWR, allowing the magnetic field intensity to be defined via a free parameter $\eta_B = P_\mathrm{mag}/P_\mathrm{th}$. Here, we introduce a slight modification f}ollowing \cite{Martinez2023}. We assume that the magnetic field decays along the CWR following the so-called frozen-in conditions in the regions where the fluid becomes transonic. This results in a faster decline of the magnetic field and, consequently, less emission being produced farther out in the CWR. This has a strong impact on the SED of HD~932129A close to its periastron, as the radio photospheres of the stellar winds cover a significant part of the brightest regions of the CWR, whereas the emission produced farther out reaches the observer unattenuated (as explained in more detail in Sect.~\ref{sec-perifromradio}). A steeper decline of the magnetic field strength along the CWR leads to a lower intrinsic luminosity in these farther out regions, {enhancing the effects of the FFA} in the integrated SED. We show this effect in the SEDs in Fig.~\ref{fig:sed_may2019} (top panel).
%\footnote{{For dense enough plasmas, there can be some significant free-free emission contribution to the thermal radio emission from the CWR, and this should more likely happen in short period systems \citep[see][]{pittard2010}. With the expected separation between the stars in HD~93129A, such a contribution is unlikely.  
%On top of that, free-free emission from the CWR would basically contribute at rather high frequencies.}}.

The SED shape is also affected by the assumed volume filling factor of the stellar winds, $f$. For simplicity we assume a constant $f$ in the winds, although it is possible that it depends on the radial distance to the star \citep{Daley-Yates2016}. A small $f$ implies a very clumpy wind, which is more efficient in both absorbing low-frequency emission and producing thermal free--free emission at higher frequencies. We explore the impact of this parameter in Fig.~\ref{fig:sed_may2019} (top panel). For $f \leq 0.6$, the SED peak lies at frequencies $\gtrsim 5$~GHz. Our radio data favours instead an SED peaking at around 3--5~GHz, requiring less absorption in the winds. For this reason, we decide to fix $f=0.9$, which is also consistent with the expected values at large radii \citep{Daley-Yates2016}. 

Finally, we can constrain {the parameter $\eta_B$ that defines the magnetic field intensity.} We first fix the fraction of the available wind kinetic power transferred to relativistic electron acceleration as $f_\mathrm{NT,e}=0.0015$ in order to be consistent with the hard X-ray fluxes derived by \cite{delPalacio2020}. We then explore the SED for different values of $\eta_B$ and present the results in Fig.~\ref{fig:sed_may2019} (bottom panel). We conclude that the best agreement is achieved for $\eta_B \approx 0.085$. This corresponds to a magnetic field strength in the apex of the CWR of $B \sim 1.1$~G. 
%These magnetic fields with surface stellar magnetic fields of ~360--400 G.

%------------------ Figure 8 -ex Fig. 7 ---
%__________________________

\subsection{Historical flux density variation}

It is clear from Fig.~\ref{fig:lightcurves}, top panel, that the flux density measured in the three ATCA bands presented here (S, C, and X) shows a decreasing trend along the interval 2019.36--2020.73 yr. The S-band values are initially higher, the decrease is more pronounced, and remains weaker during the last months of monitoring. The C- and X-band light curves start with a slight increase and show two relative maxima from Nov 2019 to Jun 2020. The behaviour of the spectral index along the same period at each band is shown in the lower panel of Fig.~\ref{fig:lightcurves}. At S-band it remains always positive, increasing between 2019.6--2020.1, and then decreasing again. In the C band, $\alpha$ starts with negative values, which then become positive around $\approx 2020$. At X-band $\alpha$ remains always negative. 
%We note that special care must be taken at the time of interpreting these figures, since the emitting regions - and the absorption effects - are not the same for the three bands. 
This behaviour is consistent with FFA playing an important role in shaping the SED, as FFA is less significant at higher frequencies.

\begin{figure}[h]
\centering
\includegraphics[width=\linewidth]{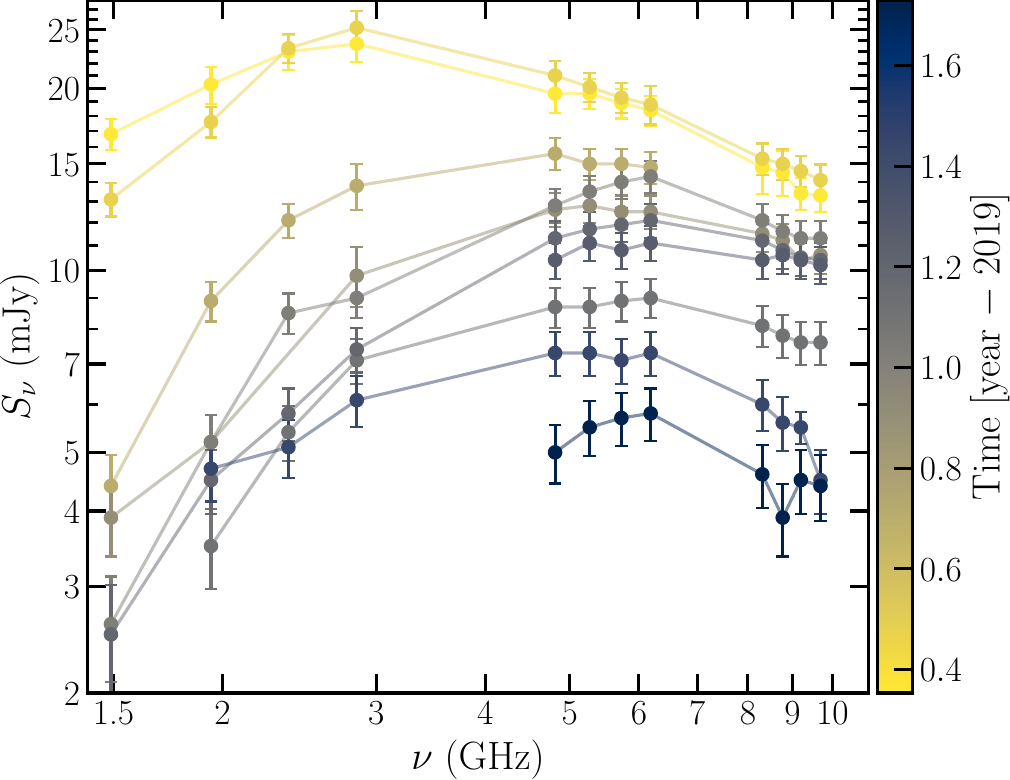}\\
\caption{Evolution of the SED of HD~93129A {observed} at the different subbands.}
\label{fig:evolution}
\end{figure}

To have a more detailed view of the evolution of the SED of HD~93129A near periastron, we have plotted it in Fig.~\ref{fig:evolution} using the totality of the flux density values in the 12 subbands from 1.4 to 10~GHz. The SED exhibits a rather smooth and uniform general trend among the subbands, with a flux density decreasing with time, but also with a turnover frequency evolving in time.  

%--------------Figure 9 - ex Figure 8 -------------
%__________________________

%-------------FIgure 10 - ex Figure 9 
\begin{figure}
\centering
\includegraphics[width=\linewidth]{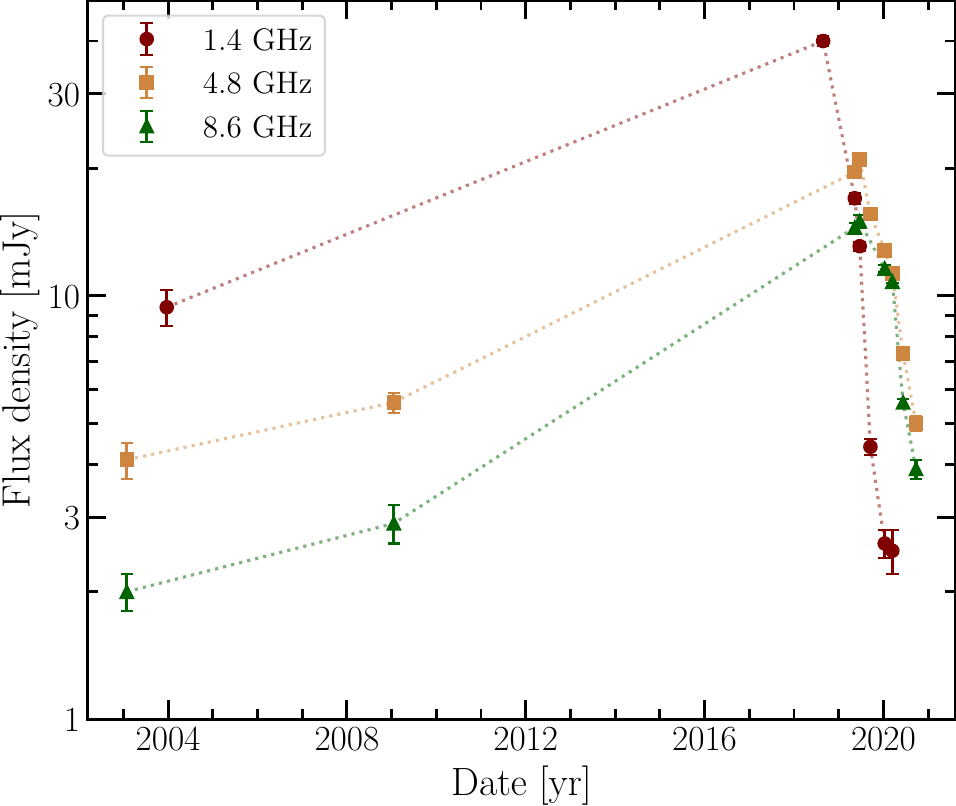} 
\caption{Light curves of HD~93129A %\textcolor{red}{\st {representing the flux densities vs time,}} 
at 1.4, 4.8 and 8.6~GHz, using data from 2003 to 2020.}
\label{fig:lc2003-2020}
\end{figure}
%__________________________

The flux densities obtained from the 2019--2020 data at subbands also allow a straightforward comparison with those at the same bands from the pre-CABB data compiled in B15 (their Table~3), at 1.4, 4.8, and 8.6~GHz. To do this, we selected from the measurements at subbands obtained in the present observations, those at S-band SB1, C-band SB1, and X-band SB2. We also added the flux density derived from the MeerKAT survey image fit {at 1.4~GHz --observation date: August 2018--} (see Fig.~\ref{fig:lc2003-2020}). 
From 2003 onwards, the emission increased until $\sim$2019,  {although there is a gap in the data flow from 2009 to 2018}. For the rest of the time, a rapid decrease took place in three bands, with particular characteristics: (i) at 1.4~GHz, the most recent measurement (Jun 2020) reached $\sim30$\% of that of Dec 2003; (ii) at 4.8~GHz, the values of Jan 2003 and Jun 2020 remained quite close (between 18\%); (iii) at 8.6~GHz, the flux density of Jun 2020 was still $\sim$50\% above that of Jan 2003.  {We note that in 2019 the spectral index between 1.4 and 4.8~GHz was positive, which suggests that the flux at 4.8 GHz was likely similar or even greater than the flux at L band in 2018. This would also suggest that the SED at C band also peaked before 2019.}

%--------------FIgure 11 - ex Figure 10 -------------
\begin{figure}[h]
\centering
\includegraphics[width=\linewidth]{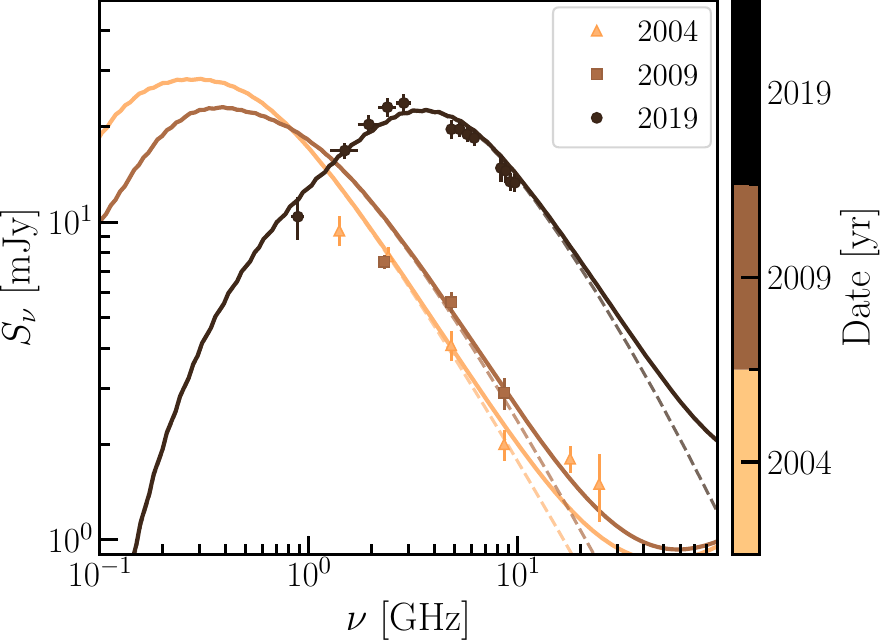}
\caption{Radio SED of HD~93129A in multiple epochs. The dashed lines {represent the synchrotron emission  
and the solid lines the total emission (including the free--free from the winds)}. } 
\label{fig:seds_multi-epoch}
\end{figure}
%__________________________

The whole dataset (observations from 2003 to 2020) allowed us to follow the evolution of the SED over a larger portion of the orbit. Figure~\ref{fig:seds_multi-epoch} shows the radio SED at three times: around 2004, around 2009, and in May 2019.  {Based on the preliminary results presented in  \cite{delPalacio2021}\footnote{ {A figure of the 3-D separation as a function of time can be accessed at \url{https://aas237-aas.ipostersessions.com/default.aspx?s=D0-A2-41-0F-11-E8-77-44-6E-67-2F-52-E6-C6-75-2C}}}, the assumed orbital separations at each epoch are 130 AU, 92 AU and 19 AU, respectively, while the projection angles $\psi$ are $83^\circ$, $75^\circ$ and $40^\circ$, based on the projected distances of $\approx$52~mas, $\approx$36~mas and 5~mas, respectively \citep{maiz2017}. We note that in \cite{delPalacio2016} an angle $\psi > 60^\circ$ was inferred for the 2009 epoch based on the morphology of the radio emission observed with VLBI \citep{benaglia2015}. For completeness, in Fig.~\ref{fig:appendix} we show how each of these parameters affects the SED.
In principle, the value of $\eta_B$ should remain constant along the orbit, given that for a stellar separation $>19$~AU, the distance from the stars to the CWR is $>50\,R_\star$, and thus the toroidal component of the stellar magnetic fields should dominate \citep[e.g.][and references therein]{delPalacio2016}. However, i}n order to reproduce the SED around 2004--2009, a smaller value of $\eta_B = 0.03$ had to be adopted (compared to $\eta_B \approx 0.085$ in May 2019). This could be indicative of magnetic field amplification close to periastron{; an alternative explanation could be that the fraction of energy going into relativistic particle acceleration does not remain constant}. The SEDs at 2004--2009 do not show  {strong} FFA at $\nu > 2$~GHz, consistent with expectations at large orbital separations.

%----------------------- Section 6 -------------
\section{Discussion on the surroundings of HD~93129A}\label{sectdiscother}

\subsection{On radio emission and spectral indices}

Figure~\ref{fig:atcafovimages} reveals the complexity of the radio continuum emission at 2.1 and 5.5~GHz in the surroundings of the colliding wind binary HD~93129A. The detected arcs, clouds and faint structures mainly come from the Carina Nebula complex (CNC). \cite{rebolledo2021} studied the different gas phases of the CNC, such as the hydrogen atomic gas  ({\sc hi}~21 cm),  the molecular gas (CO), the hot gas (8~$\mu$m) and the warm gas (70~$\mu$m). In particular, they published a detailed radio continuum map at 1--3~GHz, created using dedicated ATCA observations. They covered a $3 \times 4$~deg$^2$ region, and in their Fig.~5, they show a region that includes the one displayed by us in Fig.~\ref{fig:atcafovimages}. That is the central region of the CNC. Due to the relatively short observing times and lack of shortest baselines of the present observations, we mapped the dense cores of Car\,I and Car\,II, which host the $\eta$\,Car massive star, especially when looking at the 5.5~GHz image. Car\,I and Car\,II are two bright HII regions \citep[see][]{Brooks-2001}. The peak emission we detect at 2.1~GHz of $\eta$\,Car, Car\,I and Car\,II are  {0.4~Jy~beam$^{-1}$, 0.2~Jy~beam$^{-1}$ and 0.2~Jy~beam$^{-1}$,} respectively. These peak values are lower than those published in \cite{rebolledo2021}, presumably due to {the worse $uv$ coverage of the images presented here and consequent missing of flux.}

The spectral index distribution map shows that the presence of ionised gas is predominant in the region observed; see Fig.~\ref{spixanderror}. 
In particular, most of the ionised gas in Car\,I is optically thin except towards the edge on the east side, where the gas seems to become optically thick. The gas detected in the Car\,II region is mainly optically thick, and $\eta$\,Car emission is thermal.  {Negative spectral indices are observed towards the edge of the west side region of Car\,I,  in the region delimited by the arc structure in Car\,II and among structures between these two regions. Although there could be seen as a hint of non-thermal emission, errors in the spectral index determination are high in the places described. These regions are faint, making it difficult to measure their spectral indices accurately \citep[see, for instance,][]{green2022}, encouraging follow-up works.}
%However, both regions show negative spectral index values, revealing non-thermal emission. In the Car\,I region, this is towards the edge of the west side, while in Car\,II, it is in the region delimited by the arc structure. Nevertheless, the errors are high in the places described. Moreover, some structures in between the two regions also display non-thermal emission. It is worth noticing that the spectral index and error maps were created by using two different interferometers where the $uv$-coverage differs, and these results are triggers of follow-up works.

\subsection{Other massive binary systems in the field}

To complement our investigation, we extended the study to include the emission of other CWBs in the field (listed in Table~\ref{tab-massivestars}), excluding $\eta$\,Car, which will be the subject of a separate dedicated study. The system WR~22 is a spectroscopic binary that shows a single eclipse that occurs near the periastron passage, as the WR star passes in front of the O~star. The orbit has a period of about 80 days, as originally proposed by \citet{Moffat1978}. WR~25 shows periodic radial-velocity variations with a period of approximately 208 days \citep[][and references therein]{Gamen2006}, along with strong and variable thermal X-ray emission mainly attributable to the wind-wind interaction \citep{Arora2019}. The system HD~93250 was spatially resolved as a binary using high angular resolution observations \citep{Sana2011} and later classified as a PACWB in \citet{pacwbc} as suggested by its high radio flux density. It is made up of two very similar components, and radial velocity variations have not been detected. According to \citet{Lebouquin2017}, HD~93250 has an orbital period of around 194 days. \citet{Arora2024} reported on a clear variation in the thermal X-ray emission from HD~93250 with the same period.

We analysed the positions of these other CWBs in the field to search for radio emission. To that end, we made use of our ATCA images at 2.1, 5.5 and 9.0 GHz. 
{These images were produced to optimise the detection of point sources by keeping only the longest baselines (see Subsect.~\ref{subsecImages}).}

In the cases where we detected radio emission at the position of the systems, we measured the flux density.  {To this aim, we used the CASA task \texttt{imfit} in a region defined by the 3~$rms$ (3~$\sigma$) contour around the source. The errors were estimated 
as the quadratic sum of the $rms$ around the source, plus the fitting error resulting from \texttt{imfit}.}
For the case of non-detection, we provide an upper limit to the emission {equal to 3 times the $rms$ at the stellar system position}. The results are presented in Table~\ref{tab-maassivestarsfluxes}. 

%--------------------------Table 4------------
\begin{table}[h]
\caption{Flux densities of the massive binary systems in the field of HD~93129A in the three bands observed by ATCA.} 
\label{tab-maassivestarsfluxes}      
\centering          
\begin{tabular}{l r r r }     % 7 columns
\hline\hline 
System & $S_{\rm 2.1GHz}$ & $S_{\rm 5.5GHz}$  & $S_{\rm 9.0GHz}$ \\
            &  (mJy) & (mJy) & (mJy) \\
\hline                    
   WR~22 & $\leq 1.3 $ & out of FoV & out of FoV\\  
   WR~25 & $2.60\pm0.12$ & out of FoV & out of FoV \\
   HD~93250 & $1.98\pm0.12$ & $2.46\pm0.24$ & out of FoV\\
%   $\eta$\,Car$^e$ & LBV-like object + ? & 2.35 ?? & Tr 16 \\ 
%   WR~28$^a$  & WN6–8 + O & $6.9\pm0.5$ & \\ % MS 2
\hline                  
\end{tabular}
\end{table}
%______________________________________________

{Note that the above values were obtained from images produced from data taken over an observation timeframe of 17 months (May 2019--Sep 2020). Since the orbital periods of the three systems are of the order of the observation interval, a more appropriate approach would be to search for detections at each epoch and, if found, to measure the variation of the flux densities with epoch (eventually also subbands) to look for phase-locked imprints. This is beyond the scope of the present study and will be reported elsewhere. }

These systems were previously included in an observational campaign aimed at investigating the radio continuum emission of southern early-type stars with the ATCA telescope. The three systems were first observed in 1994, at 8.64 GHz and 4.80 GHz, by \citet{Leitherer1995}. Later, in 1997, the same team dedicated the campaign to search for radio emission from WR stars only. In \citet{Chapman1999}, they report measurements at 1.4 and 2.4 GHz from tens of southern WR stars, including WR~22 and WR~25. {In both papers, the flux density errors quoted are 1~$\sigma$ for detections and 3~$\sigma$ for upper limits, where $\sigma$ is the image $rms$ at the stellar position.}

{Our current flux density measurement for HD~93250 at 5.5~GHz ($2.46\pm0.24$~mJy, see Table~\ref{tab-maassivestarsfluxes}) 
is consistent with the upper limit that \citet{Leitherer1995} provided for the system at 4.8 GHz ($S_{\rm 4.8GHz}<3.57$ mJy). 
In the case of WR~22, \citet{Chapman1999} listed upper limits both at 2.4~GHz ($S_{\rm 2.4GHz}<0.63$ mJy) and 1.4~GHz ($S_{\rm 1.4GHz}<1.23$ mJy), consistent with the one we obtain at 2.1~GHz. 
WR~25 was detected only at 8.4~GHz ($S_{\rm 8.4GHz}=0.90\pm 0.15$ mJy) by \citet{Chapman1999}; they gave upper limits at 2.4 and 1.4~GHz ($S_{\rm 2.4GHz}<1.89$ mJy and $S_{\rm 1.4GHz}<4.11$ mJy). The difference with the conclusive detection of WR~25 presented here ($S_{\rm 2.1GHz} = 2.60\pm0.12$~mJy), especially when compared with their 2.4~GHz upper limit,} can be found in the much wider bandwidth we used (2~GHz versus 128~MHz), the longer observation time we spent on the source ($\sim$18~h vs $\sim$1~h), and that their observations were made over five days (1997 Feb 23--27), while the measurement quoted in Table~\ref{tab-maassivestarsfluxes} corresponds to observations {along the interval 2019.4--2020.7 yr.}

%----------------------- Section  7-------------

\section{Conclusions and prospects}\label{sectconcl}

The radio monitoring of HD~93129A at centimetre wavelengths with ATCA allows us to draw the following conclusions:

\begin{itemize}

\item The light curve of this extreme colliding-wind binary, between 2019 and 2020 and in the range 1.4--10~GHz, showed a general decay of $\sim$70\% in flux density.

\item When compared to historical measurements (since 2003), the peak of radio continuum emission could be established in early 2019, i.e., close to the date periastron was predicted by astrometric observations in the optical range.

\item  Preliminary SED model applied to May 2019, the date for which data including the 0.89~GHz band could be compiled, provided a first estimate of a magnetic field of around 1~G in this epoch. The fact that a significant part of the CWR is obscured by the stellar winds makes the modelling of the radio SED very sensitive to the shock conditions far from the apex.

\item The evolution of the radio SED is not consistent with a fixed value of $\eta_B$ along the orbit. This could be indicative of magnetic field amplification closer to periastron.  

\item The stronger decay of the S-band flux around $\approx$2020 might be indicative of stronger FFA effects. This could be related to orientation effects in the orbit rather than to the orbital separation, signaling the challenges of interpreting radio SEDs. 

\item The massive binary systems WR~25 and HD~93250 were detected as 2.1~GHz radio emitters.

\item Radio observations near periastron of colliding wind-driven binaries again proved to be an ideal tool to complement studies at other wavelengths to help describe the phenomenon physically. 

\end{itemize}

One of the main problems of the CWB studies was that with the classical radio interferometers, with bandwidths of MHz, the targets to be detected (and mostly as discrete sources) were scarce.  Regarding the colliding wind regions themselves, there were only a handful to map, with very long baseline interferometers. This was particularly severe for low declination sources, due to the lack of instruments in the southern hemisphere. The situation has changed dramatically in the last long decade, with the upgrade of the main instruments to GHz bandwidths -- among other improvements-- but also with the advent of the SKA pathfinders ASKAP and MeerKAT, resulting in a sharp drop in the detection threshold.
The near future is therefore very promising, with PACWBs to be confirmed, better characterised and many more to be discovered.

The radio results presented here may trigger dedicated work to advance the search for high-energy emission from HD~93129A, complementary to \citet{delPalacio2020}. Examples such as \citet{pshirkov2016} and \citet{filocomo2023} show that re-reduction of raw \textit{Fermi} data can disclose gamma-ray detections. In particular, the former study revealed the association of the CWB WR~11 with a \textit{Fermi} source {\citep{martidevesa2020}}. The second study presented evidence for the \textit{Fermi} detection of T Tauri flares by binning the data. This line of research will  contribute to increasing the number of known gamma-ray emitters associated/identified with massive binary systems, currently a group of a couple of objects. 

With a more precise characterisation of other systems like HD~93129A, of relatively long periods, which are known to overcome periastron, multi-wavelength simultaneous monitoring could be planned well in advance, from radio to gamma rays.

%-----------------------  Acknowledgements
 
\begin{acknowledgements}
{The authors are grateful to the anonymous referee for a critical reading of the manuscript and very useful suggestions. PB and JS thank the ATCA staff for help during observing.}
This research is part of the PANTERA-Stars collaboration, an initiative aimed at fostering research activities on the topic of particle acceleration associated with stellar sources\footnote{\url{https://www.astro.uliege.be/~debecker/pantera}}. This research has made use of NASA's Astrophysics Data System Bibliographic Services. This publication benefits from the support of the Wallonia-Brussels Federation of Belgium in the context of the FRIA Doctoral Grant awarded to A.~B. Blanco. 
SdP acknowledges support from ERC Advanced Grant 789410.
BM acknowledges financial support from the State Agency for Research of the Spanish Ministry of Science and Innovation, and FEDER, UE, under grant PID2022-136828NB-C41/MICIU/AEI/10.13039/501100011033, and through the Unit of Excellence Mar\'ia de Maeztu 2020--2023 award to the Institute of Cosmos Sciences (CEX2019-000918-M) and The European Research Council (ERC) under the European Union's Horizon 2020 research and innovation programme (`EuroFlash': Grant agreement No.\ 101098079).
\end{acknowledgements}

%-----------------------  The bibliography 

\bibliographystyle{aa} % style aa.bst
\bibliography{mybibliography} % your references

%----------------------------------------

\appendix

\section{{Orbital dependence of the SED}}

\begin{figure}[h]
\centering
\includegraphics[width=\linewidth]{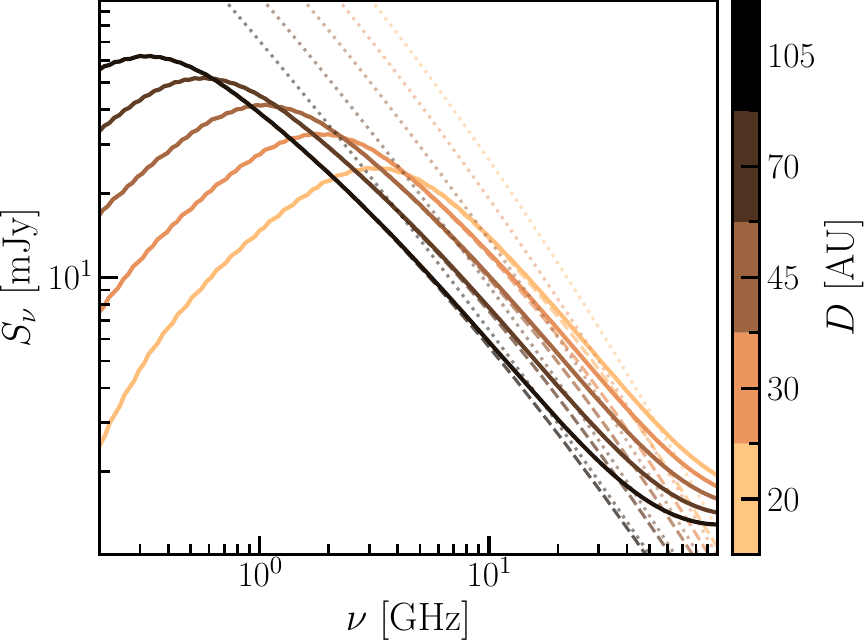}\\
\includegraphics[width=\linewidth]{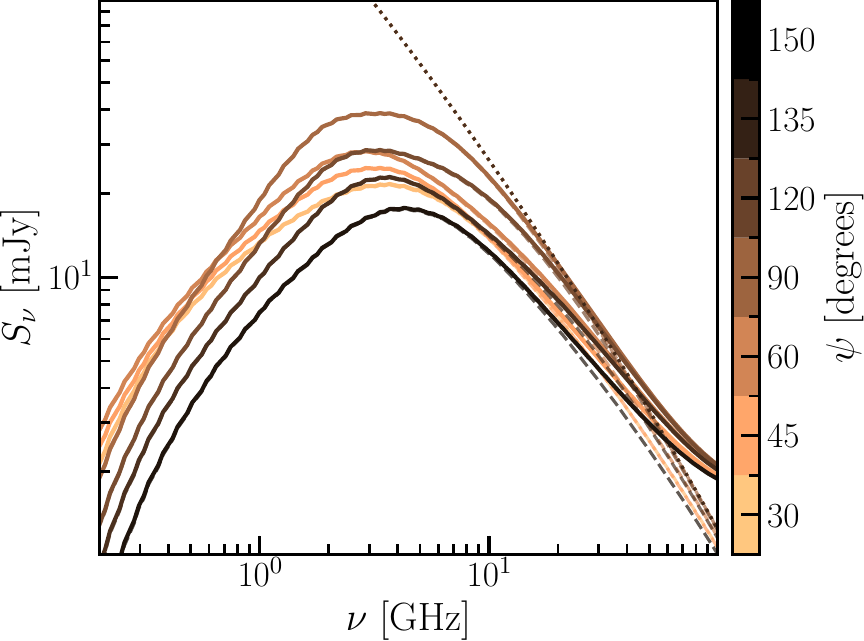}
\caption{{Radio SEDs of HD~93129A for different 3-D separations (top panel) and for different projection angles $\psi$ (bottom panel). The dotted lines represent the unabsorbed synchrotron emission (which is independent of $\psi$), the dashed lines the absorption-corrected synchrotron emission, and the solid lines the total emission (summing the contribution from the ionised winds). Values of $\psi < 90^\circ$ correspond to the secondary star being closer to the observer.}}
\label{fig:appendix}
\end{figure}

{In our emission model, the emitter is assumed to be quasi-stationary, meaning that at a given epoch the SED depends only on two orbital parameters: the 3-D orbital separation $D$ and the projection angle $\psi$ defined such that the projected orbital separation is $D_\mathrm{proj} \propto D \, \sin{\psi}$ \citep{delPalacio2022}. For completeness, in Fig.~\ref{fig:appendix} we show how these two parameters affect the SED.}

{First, we fix $\psi$ to an arbitrary value of $45^\circ$. Decreasing the parameter $D$ leads to a monotonous increase in intrinsic synchrotron emission, which enhances the emission at high frequencies. Moreover, the frequency at which FFA becomes important also increases, reducing the emission observed at lower frequencies.}

{Second, we fix $D=20$~AU and vary the parameter $\psi$. For $\psi < 90^\circ$, the secondary star is the one closest to the observer, whereas for $\psi > 90^\circ$ it is the primary that is closest. For small values of $\psi$, the radiation from the brightest portions of the CWR travel through denser parts of the wind of the secondary, and thus the absorption is strong in the SED. As $\psi$ increases to values $\sim$90$^\circ$, the absorption is minimum. For higher values of $\psi$, the emission from the CWR has to travel through denser parts of the wind of the primary, enhancing the absorption again. In this case, the absorption is even stronger than for small $\psi$-values, as the radiation has to travel through longer paths along the stellar wind before reaching the observer. This is because the primary has a stronger wind, and thus the CWR is pushed closer to the secondary star.}

{In summary, the values of $D$ and $\psi$ can have a significant impact in the SED, especially if the primary is in front. However, the SED shown in Fig.~\ref{fig:sed_may2019} is more consistent with less absorption at low frequencies, thus compatible with adopting $\psi < 90^\circ$ in Sect.~\ref{sec:sed_may2019}. Moreover, the uncertainty in $D$ is small close to periastron. Thus, we conclude that the limited knowledge of the evolution of $D$ and $\psi$ over time does not significantly affect the conclusions of our work, but it hinders our ability to provide precise measurements of the magnetic field evolution in the CWR along the orbit. Moreover, if a fast change in $\psi$ occurs close to periastron, it can already explain flux variability at frequencies $\lesssim$5~GHz within a factor $\sim$2 in this epoch.}

\end{document}